\documentclass[final,3p,times]{elsarticle}
\usepackage{psfrag}
\usepackage{mwe}
\usepackage{tikz}
\usetikzlibrary{external}
\usepackage{svg}
\usepackage{structmech}
\usepackage{float}
\usepackage{caption}
\usepackage{subcaption}
\usepackage{amsmath}

\DeclareMathOperator*{\argmin}{arg\,min}
\usepackage{comment}
\usepackage{algorithm2e}
\SetKwComment{Comment}{/* }{ */}
\usepackage{amsthm}
\usepackage{amssymb}

\usepackage{expl3}
\usepackage{enumitem}
\ExplSyntaxOn
\newcommand\latinabbrev[1]{
  \peek_meaning:NTF . {
    #1\@}%
  { \peek_catcode:NTF a {
      #1.\@ }%
    {#1.\@}}}
\ExplSyntaxOff
\usepackage{mathrsfs}

\usepackage{defwr}

\usepackage{tabularx}

\usepackage{hyperref}

\hypersetup{
    colorlinks=true,
    linkcolor=blue,
    filecolor=magenta,      
    urlcolor=cyan,
    pdftitle={Overleaf Example},
    pdfpagemode=FullScreen,
    pdfauthor=author,
    }

\usepackage{amsmath,amssymb,amsthm,color,xfrac,mathrsfs}
\usepackage{bbm}
\usepackage{makecell}
\usepackage{epsfig,amsfonts}
\usepackage{graphicx,epsfig,epstopdf} 
\usepackage{array} 
\usepackage{color,soul}
\soulregister\citep{7}
\soulregister\citep{7}
\soulregister\ref{7}
\soulregister\pageref{7}
\sethlcolor{yellow}
\usepackage{appendix}
     
\usepackage{pmat} 
\usepackage{indentfirst} 
\usepackage{placeins}
\usepackage{silence}
\usepackage{booktabs,multirow,longtable} 
\usepackage{rotating}
\usepackage{hyphenat}
\usepackage{changepage}

\NewDocumentCommand{\evalat}{sO{\big}mm}{%
  \IfBooleanTF{#1}
   {\mleft. #3 \mright|_{#4}}
   {#3#2|_{#4}}%
}

\DeclareMathAlphabet{\mathpzc}{OT1}{pzc}{m}{it}
\DeclareMathAlphabet{\mathcalligra}{OT1}{calligra}{m}{it}

\AtBeginDocument{

}

\newcolumntype{P}[1]{>{\centering\arraybackslash}m{#1}}

\graphicspath{{./Images/}}

\journal{}
\makeatletter
\def\ps@pprintTitle{%
  \let\@oddhead\@empty
  \let\@evenhead\@empty
  \let\@oddfoot\@empty
  \let\@evenfoot\@oddfoot
}
\makeatother

\begin{document}
\begin{frontmatter}

\title{A variational phase-field model for anisotropic fracture \\ accounting for multiple cohesive lengths}
\author[1,2]{Angela Maria Fajardo Lacave}
\author[2]{Francesco Vicentini}
\author[1]{Fabian Welschinger}
\author[2]{Laura De Lorenzis\corref{cor1}}
\ead{ldelorenzis@ethz.ch}

\address[1]{Robert Bosch GmbH, Corporate Sector Research and Advance Engineering,
\newline Robert Bosch Campus 1, 71272 Renningen, Germany}
\address[2]{Eidgen\"{o}ssische Technische Hochschule Z\"{u}rich, Computational Mechanics Group,
\newline Tannenstrasse 3, 8092 Z\"{u}rich, Switzerland}
\cortext[cor1]{Corresponding author}

\begin{abstract}
\label{abstract}
We propose a novel variational phase-field model for fracture in anisotropic materials. The model is specifically designed to allow a more flexible calibration of crack nucleation than existing anisotropic fracture formulations, while avoiding the introduction of multiple damage variables. In addition to the classical components of anisotropic phase-field models based on a single damage variable—namely, anisotropic elasticity and the extension of the fracture energy density via a second-order structural tensor—the proposed approach introduces fracture anisotropy through a cohesive degradation function with potentially distinct cohesive lengths along the principal material directions. For this reason, we refer to it as \textit{multi-cohesive model}. This feature enables independent control of the critical stresses governing crack nucleation in each material direction. We analyze the homogeneous solution and its second-order stability, and we compare the resulting strength surfaces with those of two representative anisotropic phase-field models available in the literature. Finally, numerical simulations in two and three dimensions demonstrate the capability of the proposed model to independently control crack nucleation and propagation in anisotropic fracture problems of increasing complexity.
\end{abstract}


\end{frontmatter}


\section{Introduction}
\label{sct:1_Introduction}

Variational phase-field models of brittle fracture, originally introduced as a regularization of the variational reformulation of Griffith's fracture theory~\cite{francfort+marigo98,Bourdin2007} and later interpreted as a family of gradient damage models \cite{Pham2011}, have become a widely used approach to study crack propagation in brittle solids. 
Classical formulations for brittle fracture typically rely on the so-called AT-1 and AT-2 models corresponding to  specific choices of the \textit{degradation} and the \textit{local dissipation} functions~\cite{Pham_2010, Ambati2014}. 
These models have been shown to asymptotically reproduce the sharp brittle fracture results in the limit of a vanishing regularization length, see e.g.~\cite{sicsic2013gradient}. 

Due to the finite regularization length, the phase-field approach is able to predict the nucleation of a new crack which, within the variational phase-field framework, is identified with the localization of the phase-field variable. In turn, adopting a stability criterion based on local minimization, localization is related to the loss of stability of the solution corresponding to an almost uniform damage value. The associated nucleation load  crucially depends on the regularization length, which leads to its interpretation as a material parameter to be calibrated based on the tensile strength of the material \cite{Tanne2018a}. The extension of this concept to fracture under multiaxial stress states, where calibration of the tensile strength is no longer sufficient to correctly reproduce experimental data, has been the subject of great attention in the recent literature, see e.g. \cite{DeLorenzis2021,vicentini2024energy,lopez2025classical}. 
To achieve a greater flexibility in the calibration of a multiaxial strength domain, a natural direction is the transition from purely brittle to cohesive models. One of the first attempts of this type has been the use of gradient damage models with more complex degradation functions, which exhibit an asymptotically cohesive behavior (see \citep{Lorentz_2010, Lorentz2012, zolesi2024stability} and the later generalization in \cite{WU201820}). A key advantage of these models is their ability to predict a tensile strength independent of the regularization length. However, in the multiaxial case, these models offer limited control over the shape of the strength surface \cite{zolesi2024stability}. Another approach, proposed in the mathematical community, consists of a family of gradient damage models that \(\Gamma\)-converge to a sharp cohesive model \cite{conti2024phase}. However, these models require modifications to make them numerically implementable \cite{freddi2017numerical}. Most recently,
a cohesive approach enabling flexibly tunable strength surfaces under multiaxial loading has been proposed in Vicentini et al.~\cite{vicentini2025variational} and Bourdin at al.~\cite{bourdin2025variational}. 

With the increasing relevance of complex materials for various applications, including polymer composites~\cite{Ernesti2020,Dean2020,Denli_2020}, ceramics~\cite{Cavuoto2024}, and biological tissues~\cite{Levy2025,Raina_2015,Gueltekin2019}, the extension of phase-field models to account for anisotropy has gained significant attention. Rooted in the microstructure of the material, anisotropy  typically leads to directional dependencies in both the elastic stiffness and the fracture toughness. 
Various modeling strategies have been developed to capture these effects. 
A first category of available models retains one damage variable, as in the isotropic case, and accounts for anisotropy through structural tensors in the non-local term of the phase-field energy functional. Some of these models focus on two-fold symmetric anisotropy through a second-order structural tensor \cite{HAKIM2009,Clayton2014,Teichtmeister2017,GULTEKIN2018,NAGARAJA2021} while others account for four-fold symmetric anisotropy via a fourth-order structural tensor and involve not only the gradient but also the second derivatives of the phase-field variable \cite{Teichtmeister2017,Li2014,Li2019_2,Nagaraja2023}. Alternatively, a second-order model able to handle four-fold symmetric anisotropy is presented in \cite{GERASIMOV2022}. In all these models, the anisotropic elastic strain energy density is degraded by a scalar-valued degradation function. However, other approaches also incorporate a decomposition of the elastic strain energy density \cite{Dijk2020,ZiaeiRad2023a}. 
A second category of anisotropic phase-field models relies on multiple damage variables associated to different damaging mechanisms \cite{Bleyer2018,Nguyen2017_2,SCHERER2022,Rezaei2022}. In such models, the elastic strain energy density is independently degraded in relation to each mechanism. Bleyer et al. \cite{Bleyer2018} rely on a degradation tensor while other approaches introduce an energy decomposition \cite{Rezaei2022}. 
While the majority of the models of both categories regularize brittle fracture, Rezaei et al. \cite{Rezaei2022} incorporate a cohesive degradation function from the family proposed by Lorentz et al. \cite{Lorentz_2010}.
 
For both categories, the focus of existing studies mostly lies on fracture propagation rather than on nucleation behavior. As shown later in this paper, models of the first category, which rely on structural tensors to induce anisotropy of the fracture properties, offer control of anisotropic crack propagation but induce only a marginal anisotropic effect in the crack nucleation behavior (hence, in the material strength). Therefore, they are not sufficiently flexible for calibration of experimental data related to strength (i.e. for specimens with no pre-existing singularities). On the other hand, models of the second category introduce anisotropic effects related to both nucleation and propagation of cracks. However, this is obtained at the price of introducing multiple damage variables, which inevitably increases the computational cost. Moreover, the drastic transition between the corresponding multiple mechanisms may be representative of some materials but not of others. 

In this paper, we propose a novel variational phase-field model for fracture of anisotropic materials, with the goal to enable a flexible calibration of anisotropic strength (related to nucleation) and toughness (controlling propagation) while keeping a single damage variable. Based on the lessons learned from phase-field modeling of isotropic fracture, we follow the strategy of transitioning from a purely brittle to a cohesive fracture model. Since this work started before finalization of the model in \cite{vicentini2025variational,bourdin2025variational}, we base our cohesive approach on the formulation in \cite{Lorentz_2010,Lorentz2012}.
In the elastic energy term of our proposed model, the anisotropic elastic energy density is degraded by a tensorial function of the damage variable, which leads to independent cohesive behaviors along the principal directions of the material. For this reason, we denote the new model as \textit{multi-cohesive}. In the fracture energy term, we follow the classical structural tensor approach and (focusing on two-fold symmetric cases) adopt a second-order structural tensor enforcing crack growth along a desired direction.
A key feature of the proposed model is the decoupling (thus the independent control) of the anisotropy related to the elastic energy and of that related to the fracture energy, which leads to the ability to independently calibrate the anisotropic fracture nucleation and propagation behavior. 
The model is discretized with the finite element method and implemented in the commercial software package Abaqus~\cite{DassaultSystemes2009} making use of the temperature-phase-field-analogy~\cite{Navidtehrani2021b}.

The paper is structured as follows: in Section~\ref{sct:2_PhaseFieldApproachForAnisotropicCohesiveFracture}, the so-called \textit{standard model} for anisotropic phase-field fracture is introduced as a natural extension of the isotropic model. The second model introduced in this section is the so-called \textit{multi-damage model} \cite{Bleyer2018}, which allows for better control over orientation-dependent critical stresses at crack nucleation through multiple phase-field variables. Finally, we propose our new \textit{multi-cohesive model} which, despite the use of a single phase-field variable, enables the independent control of the critical stresses at nucleation in the principal material directions, thereby enhancing flexibility in reproducing experimental results while retaining the computational cost of the single-variable model. 
Section~\ref{sct:3_HomogeneousState} analytically investigates the nucleation problem for the three different models, both under general multiaxial conditions and for the special case of uniaxial tension at a variable angle with respect to the principal material directions. Among other results, we analytically obtain the strength surfaces and the directional critical stresses of the three models. 
In Section~\ref{sct:LocalizedSolution}, we focus on crack propagation and investigate it numerically with the multi-cohesive model by means of several examples of increasing complexity.
Finally, we draw the main conclusions in Section~\ref{sct:Conclusion}.

As follows, we briefly summarize the notation. Vectors and second-order tensors are both denoted by boldface fonts, e.g. $\boldsymbol{u}$ and $\boldsymbol{\sigma}$ for the displacement vector and stress tensor, respectively. Fourth-order tensors are denoted by blackboard fonts, e.g. $\mathbb{C}$ for the elasticity tensor. Both second- and fourth-order tensors are underlined when expressed in Voigt notation as $\uBsig$ or $\uIC$. Moreover, for multiple damage variables $\underline{d}$ is the array that contains them. A dot denotes an inner product between equal-order tensors, whereas a colon denotes a double contraction among tensors. For a function of one variable, e.g. $g(d)$, we denote its derivative as $g'(d)$. In some cases, we abbreviate partial derivatives as follows: $P_{,d_i}(\underline{d}) = \partial P(\underline{d}) / \partial d_i.$
Given a scalar-valued function $f(x)$, we denote its negative part as
\begin{equation*}
    \langle f(x) \rangle_-=\frac{f(x)}{2} - \frac{\vert f(x) \vert}{2}.
\end{equation*}

\section{Phase-field models for anisotropic fracture}
\label{sct:2_PhaseFieldApproachForAnisotropicCohesiveFracture}

In the following, we first review two anisotropic phase-field fracture models that we take as reference in this paper, i.e. those by Teichtmeister et al.~\cite{Teichtmeister2017} and by Bleyer and Alessi~\cite{bleyer+alessi18}. We take the former one as a prototype for the models based on one damage variable, in which anisotropy is introduced by an anisotropic elastic stiffness and by a structural tensor modifying the non-local part of the fracture energy. We denote this model as \textit{standard anisotropic model} or simply \textit{standard model} (SM).  We consider the second one \cite{bleyer+alessi18} as a prototype for the models based on multiple damage variables, which associate each variable to a different damage mechanism, and we denote it as \textit{multi-damage model} (MDM). We then propose a new phase-field approach for anisotropic fracture. This model retains a single damage variable, hence it leads to the same computational effort of the SM. However, it uses a degradation function similar to the one proposed by \citep{Lorentz_2010,Lorentz2012} for the isotropic case, but with possibly distinct cohesive lengths for different material directions. For this reason, we denote it as  \textit{multi-cohesive model} (MCM). For ease of comparability, we adopt the same notation for all models.

\subsection{Energy density functional}
\label{subsct:ModelFormulations}

In the following, we present the \textit{energy density functional} $W$ for the SM, the MDM, and the novel MCM. The functional $W$ depends on the infinitesimal strain tensor $\boldsymbol{\varepsilon}(\boldsymbol{u}) = \frac{1}{2}(\nabla \boldsymbol{u} + \nabla^{T}\!\boldsymbol{u})$, where $\boldsymbol{u}$ is the displacement field, on the generalized phase-field array $\underline{d}$, and on its spatial gradient $\nabla \underline{d}$. The generalized array of phase-field variables $\underline{d}$ accounts for possibly multiple damage mechanisms at each material point; its components are the phase-field (or damage) variables $d_i \in [0,1],\,i=1,\ldots,m$ (with $m$ as the number of damage mechanisms), such that $d_i=0$ characterizes the pristine state and $d_i=1$ the fully broken material state in relation to mechanism $i$. In the case of models with one scalar damage variable (such as the SM and the MCM), $m=1$ and the generalized array simply degenerates to $d$. In general, the energy density functional is given by
\begin{equation}
W(\Bve, \underline{d}, \nabla \underline{d}) = \psi(\Bve, \underline{d}) + \gamma(\underline{d}, \nabla \underline{d}),
\label{eq:W_fun}
\end{equation}
where $\psi$ is the \textit{elastic strain energy density} and $\gamma$ is the \textit{fracture energy density}. The three models presented as follows are characterized by different expressions for these two functionals.

\subsubsection{Energy density functional of the SM}
\label{subsct:StandardAnisotropicModel}
We start with the SM~\cite{Teichtmeister2017}, which stands as a straightforward extension of isotropic phase-field brittle fracture models  towards anisotropy. 
This model introduces a single scalar variable $d$. The elastic strain energy density $\psi$ reads
\begin{equation}
\psi(\Bve,d) = \frac{1}{2} \, \Bve : \IC(d) : \Bve
\qquad\mbox{with}\qquad
\IC(d) = g(d)\,\IC_{0},
\label{eq:StrainEnergyDensity}
\end{equation}
where $\IC(d)$ is the fourth-order \textit{stiffness tensor} depending on $d$, $\IC_{0}$ is the initial (undamaged) stiffness tensor, and $g(d)$ is the so-called \textit{degradation function}. The latter is a monotonically decreasing function which satisfies the conditions $g(0)=1$, $g(1)=0$, and $g'(1)=0$; it describes the degradation of the stored elastic energy with increasing damage. Here the degradation function takes the form
\begin{equation}
g(d) = (1-d)^2
\label{eq:degradationStandardModel},
\end{equation}
which corresponds to the choice made in both AT-1 and the AT-2 models.
Using Voigt notation, indicated by underlined symbols, and restricting ourselves to orthotropic materials, the two-dimensional (2D) initial stiffness tensor takes the form
\begin{equation}
\underline{\IC}^{\scriptstyle{2D}}_{0}=
\begin{bmatrix}
C_{11} & C_{12} & 0     \\
C_{12} & C_{22} & 0     \\
0          & 0          & C_{66}
\end{bmatrix}.
\label{eq:StiffnessTensor2D}
\end{equation}
The fracture energy density $\gamma$ is written as
\begin{equation}
\label{eq:diss_en_dens_standard}
    \gamma(d,\nabla d)=\frac{\hat{G}_c}{c_w} \left( \frac{w(d)}{l}  + l \, \nabla d \cdot \boldsymbol{A} \nabla d \right), 
\end{equation}
where, as in Teichtmeister et al.~\cite{Teichtmeister2017}, anisotropy is integrated in the model response through the introduction of the second-order structural tensor $\boldsymbol{A}$. The parameter $\hat{G_{c}}$ represents a reference constant scaling the polar plot of the \textit{fracture toughness} \cite{Nagaraja2023}, $w(d)$ is the so-called \textit{local dissipation function}, a continuous and monotonically increasing function that satisfies $w(0)=0$, $w(1)=1$, and $c_{w} = 4 \int^1_0{\sqrt{w(t)}\,\textrm{d}t}$ is a normalization constant \cite{Pham_2010}. In the choice of $w(d)$ we follow the AT-1 model which leads to
\begin{equation}
w(d) = d
\qquad\mbox{and}\qquad
{c_w} = \frac{8}{3}.
\label{eq:wd_SM}
\end{equation}

\subsubsection{Energy density functional of the MDM}
\label{subsct:MultiDamageMechanismsModel}
Another phase-field model for anisotropic fracture was proposed by Bleyer and Alessi ~\cite{bleyer+alessi18} in the 2D setting. This model adopts an array of independent phase-field variables $\underline{d} = [d_1, \ldots, d_m]^\textrm{T}$ representing distinct damage mechanisms. The elastic energy density is given by
\begin{equation}
\psi(\Bve,\underline{d}) = \frac{1}{2} \, \Bve : \IC(\underline{d}) : \Bve
\qquad\mbox{with}\qquad
\IC(\underline{d}) = \ID(\underline{d}):\IC_{0}:\ID(\underline{d}).
\label{eq:StrainEnergyDensityMultipleDamage}
\end{equation}
where a fourth-order tensorial degradation function $\ID(\underline{d})$ is introduced. For 2D problems with two damage mechanisms ($m=2$), in Voigt notation, this function takes the form
\begin{equation}
\begin{array}{lclc}
\underline{\ID}^{\scriptstyle{2D}}(\underline{d}) = 
\begin{bmatrix}
\sqrt{g(d_1)} & 0 & 0 \\
0 &  \sqrt{g(d_2)} & 0 \\
0 & 0 &  \sqrt[4]{g(d_1)\,g(d_2)}\\
\end{bmatrix}.
\end{array}
\label{eq:DegradationTensorMultipleDamage}
\end{equation}
In ~\cite{bleyer+alessi18} the scalar degradation function $g(d_i)$ is chosen equal to that of the SM for all damage mechanisms
\begin{equation}
g(d_i) = (1-d_i)^2
\qquad\mbox{for}\qquad
i=1,\ldots,m.
\label{eq:g_i}
\end{equation}
The fracture energy density \( \gamma \) is given by the sum of the energy densities associated with all fracture mechanisms
\begin{equation}
    \gamma(\underline{d},\nabla\underline{d})=\sum^{m}_{i=1}\gamma_i(d_i,\nabla d_i)\qquad\mbox{with}\qquad\gamma_i(d_i,\nabla d_i)=\frac{\hat{G}_{c_i}}{c_{w_i}} \left( \frac{w(d_i)}{l_i}  + l_i \, \nabla d_i \cdot \nabla d_i \right).
\end{equation}
Bleyer and Alessi ~\cite{bleyer+alessi18} choose
\begin{equation}
w(d_i) = d_i
\qquad\mbox{and}\qquad
c_{w_i} = c_w =  \frac{8}{3}
\qquad\mbox{for}\qquad
i=1,\ldots,m.
\label{eq:wd_BA}
\end{equation}

\subsubsection{Energy density functional of the proposed MCM}
\label{subsct:ProposedModel}
We now introduce our proposed model. In contrast to the MDM, our proposed approach only accounts for a single damage mechanism, associated to a single phase-field variable $d$. The degradation is described by a fourth-order tensor as follows
\begin{equation}
    \psi(\Bve,d) = \frac{1}{2} \, \Bve : \IC(d) : \Bve
    \quad\mbox{with}\quad
    \IC(d) = \ID(d):\IC_{0}:\ID(d).
    \label{eq:StrainEnergyDensityMultipleMechanisms}
\end{equation}
where, for 2D problems, the degradation tensor $\ID(d)$ in Voigt notation reads
\begin{equation}
\underline{\ID}^{\scriptstyle{2D}}(d)=
\begin{bmatrix}
\sqrt{g_1(d)} & 0 & 0 \\
0 &  \sqrt{g_2(d)} & 0 \\
0 & 0  &  \sqrt[4]{g_1(d)\,g_2(d)} \\
\end{bmatrix}
\label{eq:DegradationTensorProposedModel_2D}
\end{equation}
and the scalar degradation functions $g_i(d),\,i=1,2$, are 
chosen as \citep{Lorentz_2010, Lorentz2012} 
\begin{equation}
g_i(d) = \frac{\,(1-d)^{2}}{(1-d)^{2} + 2\,r_i\,d\,(1+p_i\,d)}\qquad \text{with} \qquad r_i = \frac{\ell_{c_i}}{\ell}, \qquad i=1,2.
\label{eq:degradationLorentz}
\end{equation}
These functions are continuously differentiable for $d \in [0,1]$ and fulfill the conditions $g_i(0) =1$, $g_i(1) = 0$, and $g_i'(d) < 0$. The latter condition yields the admissible range for the cohesive parameters $p_i > -1$ and $r_i > c_w\,(2+p_i)/2$. In the following, for simplicity we always assume $p_1=p_2=p$.
For an extensive discussion of the cohesive degradation law (\ref{eq:degradationLorentz}) see Lorentz~\cite{Lorentz2012}.

Note the similarity between the degradation tensor of the MDM in equation~(\ref{eq:DegradationTensorMultipleDamage}) and the degradation tensor of the MCM in equation~(\ref{eq:DegradationTensorProposedModel_2D}). In the former, anisotropic degradation effects of the elastic strain energy density are accounted for via multiple  damage mechanisms, each described by a phase-field variable $d_i$, and a single degradation function $g(d_i),\,i=1,2$. In the latter, anisotropic degradation is realized with a single phase-field variable $d$ but with different  degradation functions $g_i(d),\,i=1,2$. 

As in the SM, see Section~\ref{subsct:StandardAnisotropicModel}, the fracture energy density is given by~\eqref{eq:diss_en_dens_standard}, i.e. anisotropy is incorporated via the structural tensor~$\BA$ in the non-local term of the fracture energy density. In the choice of $w(d)$ we follow the AT-1 model, hence \eqref{eq:wd_SM} holds.

\subsection{Time-discrete variational formulation}
\label{subsct:VariationalFormulation}
This subsection outlines the time-discrete variational formulation yielding the governing equations for the three models presented earlier. To obtain the solution within a typical time increment $[t_n, t_{n+1}]$, we assume all variables at time $t_n$ to be known and compute the solution fields at time $t_{n+1}$ based on a time-discrete variational principle. For the sake of a compact notation, we drop the subscript $n+1$ and assume all variables without subscript to be evaluated at time $t_{n+1}$. 
The phase-field fracture problem is governed by the \textit{total energy} functional
\begin{equation}
\mathcal{E}(\Bu,\underline{d}) = \!\!\int_{\Omega} W(\Bve(\Bu),\underline{d},\!\nabla \underline{d}) \, d\Omega - \!\int_{\Omega} \Bf \cdot \Bu \, d\Omega  - \!\int_{\partial\Omega_t} \Bt_{\text{\scriptsize{N}}} \cdot \Bu \, d\Gamma, 
\label{eq:VariationalProblem}
\end{equation}
where $\Omega \subset \mathbb{R}^{n_{dim}}$ is the domain of interest (with $n_{dim}$ as the number of space dimensions) and $\Bf$ are the prescribed body forces at time $t_{n+1}$. The domain boundary $\partial\Omega$ is decomposed into the Dirichlet part $\partial\Omega_u$, with prescribed displacements $\Bu_D$,
and the (complementary) Neumann part $\partial\Omega_t$, with prescribed tractions $\Bt_{\text{\scriptsize{N}}}$. Both Dirichlet and Neumann boundary conditions in general depend on time and $\Bu_D$ and $\Bt_N$ denote the values at time $t_{n+1}$.
The field variables at time $t_{n+1}$ are obtained by solving the variational problem
\begin{equation}
    (\Bu,\underline{d}) = \mathop{\arg \text{loc} \min}\limits_{(\Bu^\ast, \underline{d}^\ast) \ \in \ \mathcal{C} \times \mathcal{D}_n}  \mathcal{E}(\Bu^\ast, \underline{d}^\ast)
    \label{eq:arglocmin}
\end{equation}
where 
\begin{equation}
\begin{split}
& \mathcal{C} = \left\{ \boldsymbol{u} \in H^1(\Omega;\mathbb{R}^{n_{dim}}): \boldsymbol{u} = \boldsymbol{u}_D \ \text{on}\ \partial\Omega_u \right\}
\qquad\mbox{and}\qquad 
\mathcal{D}_n = \left\{\underline{d}: d_i \in H^1(\Omega;\mathbb{R}),\,d_i\geq d_{in},\, i=1,\ldots,m\right\}
\end{split}
\label{spaces}
\end{equation}
are the spaces of the admissible displacement and damage fields at time $t_{n+1}$. In (\ref{spaces}) $d_{in}$ is the value of the $i$-th damage variable at the previous time step $t_n$. 
The local constrained minimization problem \eqref{eq:arglocmin} yields the following first-order stability condition
\begin{equation}
\delta \mathcal{E}(\Bu,\underline{d})(\Bu^\ast-\Bu, \underline{d}^\ast-\underline{d}) \geq 0 \quad \forall (\Bu^\ast,\underline{d}^\ast) \in \ \mathcal{C} \times \mathcal{D}_n
\label{eq:NecessaryCondition}
\end{equation}
where
\begin{equation}
\delta\mathcal{E}(\Bu,\underline{d})(\delta\Bu,\delta\underline{d}) = \frac{d}{dh}\mathcal{E}(\Bu + h\,\delta\Bu, \underline{d} + h\,\delta\underline{d}) \Bigg\vert_{h=0}
\end{equation}
is the \textit{Gâteaux} derivative of $\mathcal{E}$ in the admissible direction $(\delta\Bu, \delta\underline{d})$. 
For smooth solutions, standard arguments of calculus of variation deliver the strong form of the governing equations.
The variation with respect to the displacement field gives the equilibrium equation and the natural boundary conditions
\begin{equation}
\operatorname{div}\,\Bsigma(\Bve,\underline{d}) + \Bf = \boldsymbol{0} \quad\text{in}\quad \Omega
\qquad\mbox{and}\qquad
\Bsigma(\Bve,\underline{d})\cdot\Bn = \Bt_N \quad\text{on}\quad \partial\Omega_t
\label{eq:Partial_Psi_u}
\end{equation}
where we introduced the stress
\begin{equation}
\Bsigma(\Bve,\underline{d}) = \dfrac{\partial \psi(\Bve,\underline{d})}{\partial \Bve} 
= \IC(\underline{d}) : \Bve
\label{eq:sigma_definition}
\end{equation}
with the damaged stiffness tensor $\IC(\underline{d})$ taking different representations for the individual models. The variation of the total energy functional with respect to the damage variables yields the Karush–Kuhn–Tucker (KKT) conditions
\begin{equation}
G_i(\Bve,\underline{d}) \leq R_i(d_i,\nabla d_i),\qquad
d_i \ge d_{in},\qquad
[-G_i(\Bve,\underline{d})+R_i(d_i,\nabla d_i)]\,(d_i-d_{in})=0
\quad\mbox{for}\quad
i=1,\ldots,m
\quad\mbox{in}\quad
\Omega
\label{eq:damageCriterion}
\end{equation}
with $m=1$ for SM and MCM. The KKT conditions~(\ref{eq:damageCriterion}) consist of (i) the damage criteria, which we express in terms of the \emph{damage energy release rates} $G_i(\Bve,\underline{d})$ and of the \emph{damage resistances} $R_i(d_i,\nabla d_i)$
\begin{equation}
G_i(\Bve,\underline{d}) = -\dfrac{\partial \psi(\Bve,\underline{d})}{\partial d_i}
\qquad\mbox{and}\qquad
R_i(d_i,\nabla d_i) = \dfrac{\partial \gamma_i(d_i,\nabla d_i)}{\partial d_i} - \nabla \!\cdot\! \dfrac{\partial \gamma_i(d_i,\nabla d_i)}{\partial (\nabla d_i)},
\label{eq:GiRi_defs}
\end{equation}
(ii) the irreversibility constraints, and (iii) the consistency conditions. We also obtain, again in KKT form, the natural boundary conditions for the damage fields
\begin{equation}
\nabla d_i \cdot \Bn \geq 0,
\qquad
d_i \geq d_{in},\qquad
[\nabla d_i \cdot \Bn]\,(d_i - d_{in}) = 0
\quad\mbox{for}\quad
i=1,\ldots,m
\quad\mbox{on}\quad
\partial\Omega
\label{eq:XXX}
\end{equation}
with $m=1$ for SM and MCM. 

\subsection{Irreversibility and non-negativity constraints}
\label{subsubsct:IrreversibiltyAndNon-NegativeConstraints}
In an algorithmic setting, the KKT conditions~(\ref{eq:damageCriterion}) and \eqref{eq:XXX} can be handled in various ways. Here, we enforce irreversibility in the evolution of the phase-field variables in an approximate fashion by augmenting the energy (\ref{eq:VariationalProblem}) with a penalty term as follows
\begin{equation}
\calE^p(\Bu,\underline{d}) = \calE(\Bu,\underline{d}) + P(\underline{d})
\qquad\mbox{with}\qquad
P(\underline{d}) = \sum_{i}^m \frac{\lambda_i}{2} \langle d_i - {d_i}_n \rangle_{-}^2 + \sum_{i}^m \frac{\rho_i}{2} \langle d_i \rangle_{-}^2.
\label{eq:PenaltyTerm}
\end{equation}
The second term in $P(\underline{d})$, needed in case of initial cracks to be described by their phase-field regularization, enforces the non-negativity of the phase-field variables. An estimate for a good choice of the penalty parameters $\lambda_i$ and $\rho_i$ can be found in \cite{Gerasimov2019}. In this way, the KKT conditions \eqref{eq:damageCriterion} and
\eqref{eq:XXX} are replaced by the equalities
\begin{equation}
G_i(\Bve,\underline{d}) - R_i(d_i,\nabla d_i) - P_{,d_i}(\underline{d}) = 0 \quad\text{in}\quad \Omega
\qquad\mbox{and}\qquad
\nabla d_i \cdot \Bn = 0 \quad\text{on}\quad \partial\Omega
\label{eq:Partial_Psi_d_penalty}
\end{equation}
Thus, \eqref{eq:Partial_Psi_u} and \eqref{eq:Partial_Psi_d_penalty} give a set of coupled equations (and Neumann boundary conditions) in the displacement field $\Bu$ and the  phase-field variable(s) $\underline{d}$.

\subsection{Admissible strain and stress domains, strength surfaces}
\label{subsubsct:local_model}
An important notion for the following analysis of damage nucleation under multiaxial stress states is that of \textit{admissible strain and stress domains}. In the context of local damage modeling, these are the sets in which, respectively, strains and stresses must remain in order for the material to follow a linearly elastic behavior without damage evolution, i.e., $d_i=d_{in} \, \forall i=1,\dots,m$. In our non-local damage modeling context, their boundaries define the elastic limits for materials with homogeneous damage distribution ($\Delta d_i=0 \, \forall i=1,\dots,m$), or at the level of a material point. For the three models that we introduced, the \textit{admissible strain domains} $\calR_i(\underline{d})$ and the \textit{admissible stress domains} $\calS_i(\underline{d})$ associated to damage mechanisms $i=1,\ldots,m$ read
\begin{align}
\mathcal{R}_i(\underline{d}) &= \left\{ \Bve \in \mathrm{Sym} \; : \;
-\frac{\partial \psi(\Bve,\underline{d})}{\partial d_i}
\leq \frac{\partial \gamma_i(d_i,\mathbf{0})}{\partial d_i} \right\}, \quad i=1,\ldots, m, \label{1}\\[6pt]
\mathcal{S}_i(\underline{d}) &= \left\{ \Bsigma \in \mathrm{Sym} \; : \;
\frac{\partial \psi^\star(\Bsigma,\underline{d})}{\partial d_i}
\leq \frac{\partial \gamma_i(d_i,\mathbf{0})}{\partial d_i} \right\},\quad i=1,\ldots,m,\label{2}
\end{align}
where Sym denotes the space of symmetric tensors, and with $m=1$ for the SM and the MCM. For the admissible stress domain, we introduced the \textit{complementary energy} $\psi^\star$
\begin{equation}
\psi^{\star}(\Bsigma,\underline{d}) = \sup_{\Bve \,\in\, \mathrm{Sym}} \left[ \Bsigma : \Bve - \psi(\Bve,\underline{d}) \right]
\end{equation}

As discussed in \cite{Pham_2010}, it is important for the local damage model to satisfy two properties denoted as \textit{strain hardening} and \textit{stress softening}. With respect to the damage variable $d_i$, the former property implies that $\mathcal{R}^{(1)}_i \subset \mathcal{R}^{(2)}_i$ if $d^{(1)}_i < d^{(2)}_i$ and guarantees a unique solution for $d_i$ at given strain, whereas the latter property, which must hold at least for a sufficiently large value of $d_i$ and is important to obtain localization of $d_i$, implies that $\mathcal{S}^{(1)}_i \supset \mathcal{S}^{(2)}_i$ if $d^{(1)}_i < d^{(2)}_i$.

For brittle isotropic materials it is well known that, for a sufficiently large structure, a necessary and sufficient condition for the damage localization in cracks is the transition from a stress-hardening to a stress-softening phase, which for the AT$_1$ model corresponds to the elastic limit at $d=0$ \citep{Pham_2010, Pham2011a,pham2013stability}. For different types of phase-field models, these conclusions do not necessarily hold and a more careful study is necessary, see also \citep{pham2013stability,zolesi2024stability}. To the best of our knowledge, second-order stability has not yet been investigated for the anisotropic case. Moreover, even for isotropic materials no second-order stability analyses are available for the asymptotically cohesive degradation function \eqref{eq:degradationLorentz}. In \ref{sct:Appendix} and \ref{AppB}, we address these gaps by performing a second-order stability analysis: in \ref{sct:Appendix}, we apply the results of \cite{Pham_2010, Pham2011} to the one-dimensional (1D) model with the degradation function \eqref{eq:degradationLorentz}, while in \ref{AppB}, we apply the analytical findings of \cite{pham2013stability} to the proposed MCM. For these studies, we use the AT$_1$ local dissipation function \eqref{eq:wd_SM}. The analysis reveals distinct behaviors between 1D and higher-dimensional settings. In 1D, the homogeneous solution becomes unstable for a sufficiently large bar, i.e., when \(L / \ell \to \infty\), where $L$ denotes the characteristic size of the domain. In contrast, in 2D, this condition alone is not sufficient to determine stability, which instead depends on the tensor \(\boldsymbol{q}(d) = \mathbb{S}'(d) \boldsymbol{\sigma}\) and on the ratios between the cohesive and regularization lengths, \(r_1\) and \(r_2\). Stability is generally favored by  distance between \(\boldsymbol{q}(d)\) and its projection onto the jump compatibility space defined in \eqref{eq:jump_compatible_space}. In 2D, a larger distance corresponds to the predominance of the volumetric component of \(\boldsymbol{q}(d)\) over its deviatoric component \cite{zolesi2024stability}. For brittle isotropic materials, such predominance arises when the applied homogeneous stress \(\bar{\boldsymbol{\sigma}}\) is itself dominated by its volumetric part \cite{zolesi2024stability}.
For the anisotropic MCM, the behavior is more intricate: under purely volumetric stress states, anisotropy tends to reduce the volumetric contribution in \(\boldsymbol{q}(d)\), thereby promoting instability of the homogeneous solution (see \ref{AppB_vol}); conversely, under uniaxial stress states, where the volumetric component of \(\bar{\boldsymbol{\sigma}}\) is limited compared to the deviatoric one, anisotropy may enhance the volumetric part of \(\boldsymbol{q}(d)\) and stabilize the homogeneous solution (see \ref{AppB_uni}).
Finally, we conclude that increasing the parameters \(r_1\) and \(r_2\) promotes the stability of the homogeneous solution, which is consistent with previous observations for other phase-field models of cohesive fracture \cite{zolesi2024stability, vicentini2025variational}. 

For the SM and the MDM, we do not perform specific investigations on second-order stability but we rely on previous numerical evidence, although partial ~\cite{Teichtmeister2017, Bleyer2018}. Based on the above, assuming to be under conditions which lead to an unstable homogeneous solution at the onset of damage, in the following we associate the localization of the damage variable $d_i$ to the attainment of the boundary of the admissible stress domain, $\partial \calS_i$, for $d_i=0$ (with all other variables also equal to zero, i.e. $\underline{d}=\underline{0}$). Thus,  $\partial \calS_i(\underline{0})$, represents the \textit{strength surface} associated to damage mechanism $i$, whereby

\begin{equation}
\partial\mathcal{S}_i(\underline{0}) = \left\{ \Bsigma \in \mathrm{Sym} \; : \;
\frac{\partial \psi^\star(\Bsigma,\underline{0})}{\partial d_i}
= \frac{\partial \gamma_i(0,\mathbf{0})}{\partial d_i} \right\},\quad i=1,\ldots,m,\label{2a}
\end{equation}
with $m=1$ for the SM and the MCM.
\subsection{Directional critical stress under uniaxial conditions}
\label{directional}
A particularly simple but interesting setting is a state of uniaxial tensile stress in two dimensions, whereby the uniaxial stress is applied at a variable angle $\theta$ with respect to the $1$-direction of the material. Under these conditions, the stress tensor is given by
\begin{equation}
\label{1D_rotated}
\tilde{\boldsymbol{\sigma}}=\sigma\left[\begin{array}{cc}
\cos^{2}\theta & -\sin\theta\cos\theta\\
-\sin\theta\cos\theta & \sin^{2}\theta
\end{array}\right]
\end{equation}
where $\sigma >0$ is the applied stress value. We define the \textit{directional critical stress} associated to the damage mechanism $i$ as the value $\sigma^{cr}_{i} >0$ for which

\begin{equation}
\frac{\partial \psi^\star(\tilde{\Bsigma},\underline{0})}{\partial d_i}
= \frac{\partial \gamma_i(0,\mathbf{0})}{\partial d_i},\quad i=1,\ldots,m.\label{2bis}
\end{equation}

\section{Strength surfaces and directional critical stresses of phase-field models for anisotropic fracture}
\label{sct:3_HomogeneousState}
\begin{figure}[h!]
    \centering
    \includegraphics[height=0.35\textwidth]{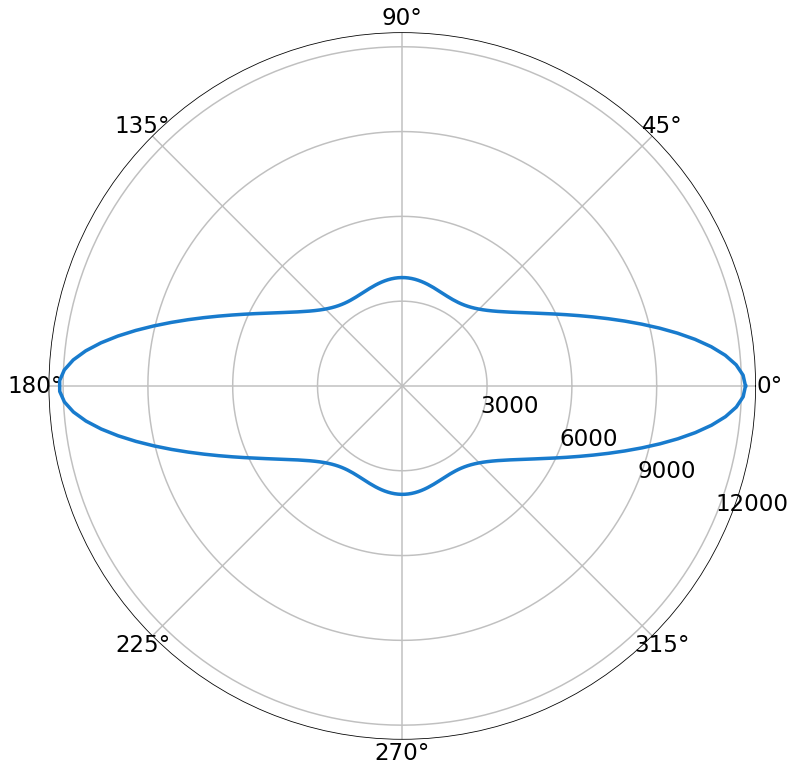}
   \caption{Orientation-dependent Young's modulus for 2D analysis.}
    \label{Figure:Emoduli2D}
\end{figure}

In the following, we consider the orthotropic 2D case and derive for the individual phase-field models in Section \ref{sct:2_PhaseFieldApproachForAnisotropicCohesiveFracture} the admissible strain and stress domains and the \textit{strength surfaces}. Moreover, we consider the special case of a uniaxial tensile stress and derive the \textit{directional critical stresses} of the three models. To illustrate the results, we assume the model parameters in Table~\ref{tab:model_params_2D}, which are realistic for a polymer-based composite reinforced with short fibers with a high degree of fiber alignment along the $1$-direction. In the uniaxial setting of Section \ref{directional}, this corresponds to the orientation-dependent Young's modulus in Figure \ref{Figure:Emoduli2D}.
\begin{table}[H]
  \caption{Material parameters used for the 2D nucleation study in Section~\ref{sct:3_HomogeneousState}.}
  \label{tab:model_params_2D}
  \centering
  \scriptsize
  \renewcommand{\arraystretch}{0.9}
  \setlength{\tabcolsep}{4pt}
  \begin{tabularx}{\textwidth}{@{}p{0.25\textwidth}>{\centering\arraybackslash}X >{\centering\arraybackslash}X >{\centering\arraybackslash}X@{}}
    \toprule
    \textbf{Parameter} &
    \multicolumn{3}{@{}l@{}}{\textbf{All models}}
    \\
    \midrule
    Elastic coefficients (N/mm$^2$) &
    \multicolumn{3}{@{}l@{}}{
    $
    \setlength\arraycolsep{0pt}
    \begin{array}{ll}
        C_{11}=14945            & \quad C_{12}=3970 \\
        C_{22}=\phantom{1}6582  & \quad C_{66}=1295
      \end{array}
    $
    }
    \\
    \midrule
    &
    \textbf{SM} &
    \textbf{MDM} &
    \textbf{MCM} \\
    \midrule
    Regularization lengths (mm) &
    $\ell=1.0$ &
    $\ell_1=\ell_2=1.0$ &
    $\ell=1.0$ \\
    Cohesive lengths (mm) &
    -- &
    -- &
    $\ell_{c_i} = [1.0, \ldots, 6.0]$ \\
    Cohesive degradation parameter (–) &
    -- &
    -- &
    $p = 2.0$ \\
    Fracture energy scaling factors (N/mm) &
    $\hat{G}_{c} = [1.0,\ldots, 0.1]$ &
    $\hat{G}_{c_i} = [1.0,\ldots, 0.1]$ &
    $\hat{G}_{c} = 1.0$ \\
    \bottomrule
  \end{tabularx}
  \vspace{4pt}
\end{table}

\subsection{Standard model}
\label{subsubsct:GoverningEqStandardModel}
As follows, we derive and illustrate the strength surface and the directional critical stress of the SM presented in Section \ref{subsct:StandardAnisotropicModel}.
\subsubsection{Strength surface of the SM}
For the SM, the damage energy release rate $G(\Bve,d)$ and the local resistance $R(d,\mathbf{0})$ read
\begin{equation}
G(\Bve,d) = -\frac{1}{2}\, \boldsymbol{\varepsilon} : \IC_{,d}(d) : \boldsymbol{\varepsilon}
\qquad\mbox{and}\qquad
R(d,\mathbf{0}) = \frac{\hat{G}_c}{c_w \ell}
\end{equation}
where $\IC_{,d}(d) = g'(d)\,\IC_{0} 
= -2(1-d)\,\IC_{0}$.
Thus, the admissible strain and stress domains \eqref{1},\eqref{2} take the form
\begin{equation}
\calR(d) = \left\{ \Bve \in \mathrm{Sym}: \quad
\Bve : \IC_{0} : \Bve \;\leq\;
\frac{\hat{G}_c}{c_w \ell}\,\frac{1}{1-d} \right\},
\label{eq:StrainHardeningDomain_SM}
\end{equation}
\begin{equation}
\calS(d) = \left\{ \Bsigma \in \mathrm{Sym}: \quad
\Bsigma : \IS_{0} : \Bsigma \;\leq\; \frac{\hat{G}_c}{c_w \ell}\,(1-d)^3 \right\},
\label{eq:StressSofteningDomain_SM}
\end{equation}
with the initial (undamaged) compliance tensor $\IS_{0} = \IC_0^{-1}$. These domains  evidently satisfy both the strain hardening and the stress softening conditions for all values of $d\in [0,1)$.
The strength surface follows as
\begin{equation}
\partial \calS(0) = 
\left\{ \boldsymbol{\sigma} \in \mathrm{Sym}: \;\; \boldsymbol{\sigma} : \IS_0 : \boldsymbol{\sigma}
= \frac{\hat{G}_c}{c_w \ell} \right\}
\label{eq:DamageNucleationSurface_SM}
\end{equation}

For the orthotropic 2D case in Voigt notation, with \eqref{eq:StiffnessTensor2D} and the initial compliance
\begin{equation}
\underline{\IS}^{\scriptstyle{2D}}_{0}=
\begin{bmatrix}
S_{11} & S_{12} & 0     \\
S_{12} & S_{22} & 0     \\
0      & 0      & S_{66}
\end{bmatrix},
\label{eq:DegradatedStiffnessStandardModel}
\end{equation}
the admissible domains in (\ref{eq:StrainHardeningDomain_SM}),(\ref{eq:StressSofteningDomain_SM}) reduce to
\begin{align}
(1-d)\left[C_{11}\varepsilon_{11}^2 + 2C_{12}\varepsilon_{11}\varepsilon_{22}
+ C_{22}\varepsilon_{22}^2 + 4C_{66}\varepsilon_{12}^2 \right]
&\;\leq\; \frac{\hat{G}_c}{c_w\ell}, \\[6pt]
\frac{1}{(1-d)^3}\left[ S_{11}\sigma_{11}^2 + 2S_{12}\sigma_{11}\sigma_{22}
+ S_{22}\sigma_{22}^2 + S_{66}\sigma_{12}^2 \right]
&\;\leq\; \frac{\hat{G}_c}{c_w\ell},
\label{eq:ElasticDomains_SM}
\end{align}
and the strength surface reads
\begin{equation}
\label{Strength_SM}
S_{11}\sigma_{11}^2 + 2S_{12}\sigma_{11}\sigma_{22} + S_{22}\sigma_{22}^2 + S_{66}\sigma_{12}^2
= \frac{\hat{G}_c}{c_w \ell}.
\end{equation}
From this expression, we can easily obtain the values of the critical stress components as follows
\begin{flalign}
\begin{array}{@{}l @{\qquad} l@{}}
\sigma_{11}^{\scriptstyle{cr}}=\sqrt{\dfrac{\hat{G}_c}{c_w\,\ell\,S_{11}}} &\quad\mbox{if}\quad \sigma_{22}=\sigma_{12}=0, \\[4mm]
\sigma_{22}^{\scriptstyle{cr}}=\sqrt{\dfrac{\hat{G}_c}{c_w\,\ell\,S_{22}}} &\quad\mbox{if}\quad \sigma_{11}=\sigma_{12}=0, \\[4mm]
\sigma_{12}^{\scriptstyle{cr}} =\sqrt{\dfrac{\hat{G}_c}{c_w\,\ell\,S_{66}}} &\quad\mbox{if}\quad \sigma_{11}=\sigma_{22}=0.
\end{array}
\label{eq:sigma_cr_axis_SM}
\end{flalign}
Figure \ref{subfig:StrengthSurfaces_SM} illustrates the strength surface \eqref{Strength_SM} in the $\sigma_{11}-\sigma_{22}$ plane assuming $\sigma_{12}=0$. A modification of the $\hat{G}_{c}/\ell$ ratio leads to a homothetic transformation of the strength surface, whereas its shape is only controlled by the ratios of the anisotropic elastic coefficients. Analogously, \eqref{eq:sigma_cr_axis_SM} shows that the ratios among different critical stress components only depend on the ratios of the relevant elastic compliance values.
\subsubsection{Directional critical stress of the SM}
\begin{figure}[H]
  \centering
  \begin{subfigure}{0.44\textwidth}
    \centering
    \includegraphics[width=\textwidth]{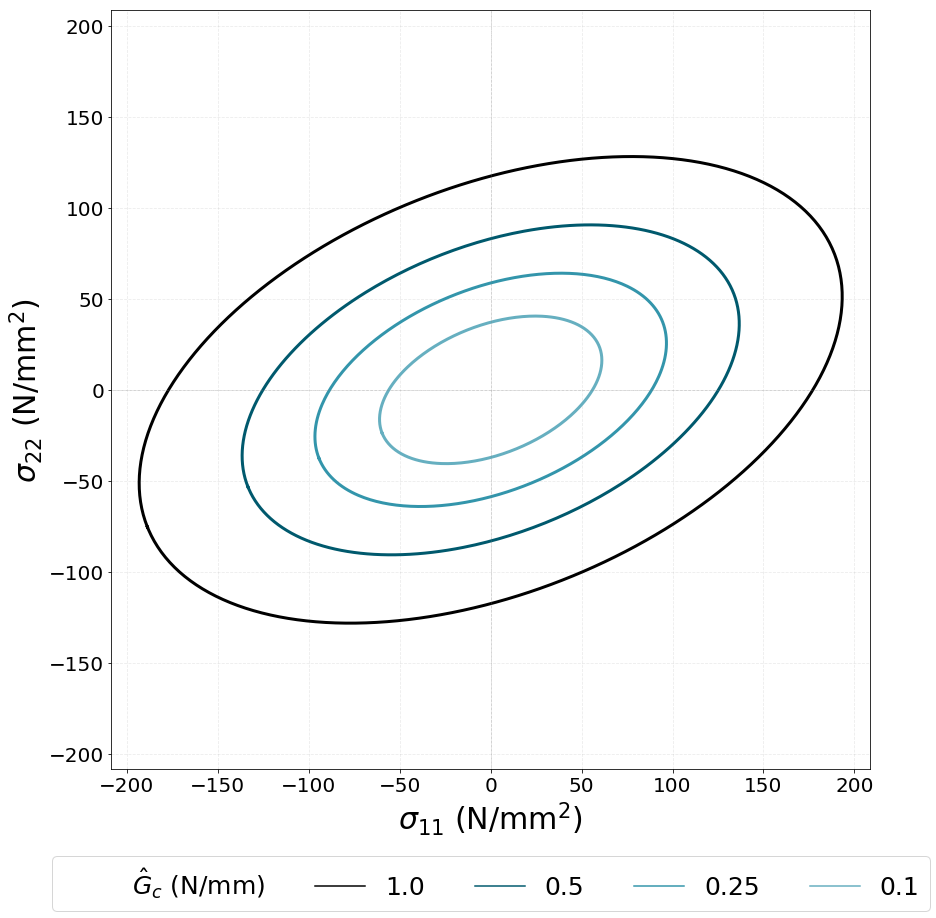}
    \caption{Strength surfaces of the SM for varying  $\hat{G}_c$ (assuming $\sigma_{12}=0$).}
    \label{subfig:StrengthSurfaces_SM}
  \end{subfigure}
   \hfill
  \begin{subfigure}{0.42\textwidth}
    \centering
    \includegraphics[width=\textwidth]{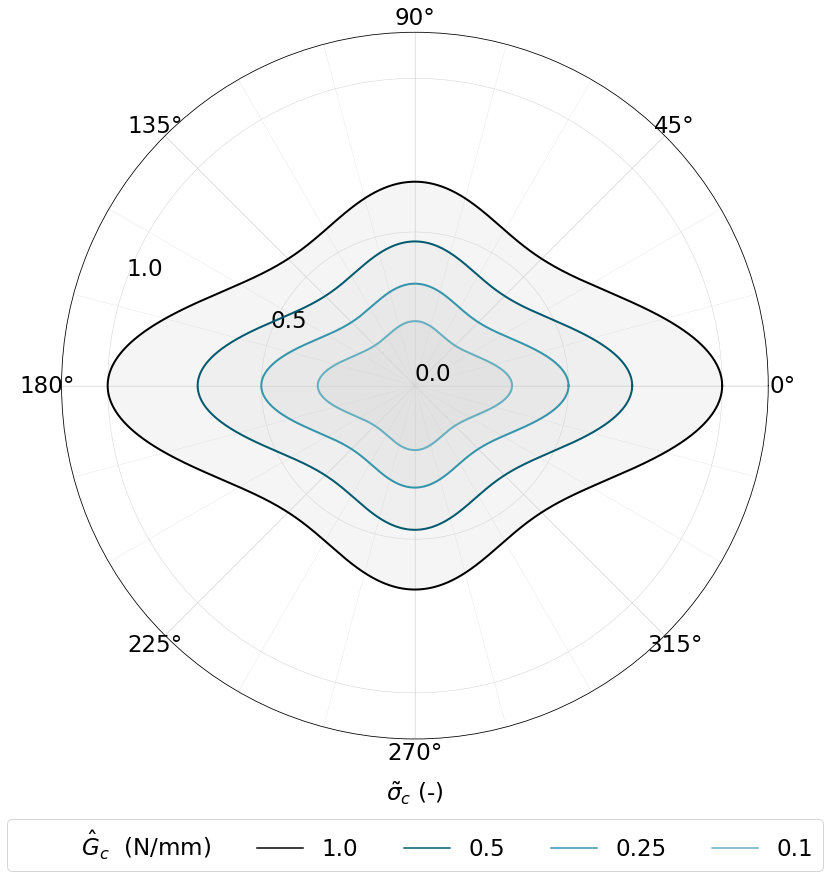}
    \caption{Polar plot of the normalized directional critical stress $\tilde{\sigma}^{\scriptstyle{cr}}(\theta) \!=\! \sigma^{\scriptstyle{cr}}(\theta) / \sigma^{\scriptstyle{cr}}(\theta \!=\! 0^\circ, \hat{G}_{c} \!\!=\! 1.0 \, \mbox{N/mm})$ for varying $\hat{G}_c$.}
    \label{subfig:HomogeneousSolution_SigmaCR_Polar_SM}
  \end{subfigure}
  \caption{Strength surface and polar plot of the directional critical stress according to the \textit{SM}}
  \label{Figure:HomogeneousSolution_SM}
\end{figure}

From the strength surface \eqref{Strength_SM}, by substituting the rotated uniaxial stress state \eqref{1D_rotated} we easily derive the directional critical stress 
\begin{equation}
    \sigma^{\scriptstyle{cr}}(\theta) = \sqrt{\dfrac{\hat{G}_c}{c_w \ell S(\theta)}}
\label{eq:CriticalStressStandardModel_simplified}
\end{equation}
where
\begin{equation}
    S(\theta) =
    S_{11}\cos^4\theta + S_{22}\sin^4\theta + \left(2S_{12} + S_{66}\right)\sin^2\theta\cos^2\theta.
\label{eq:CriticalStressSM_simplified}
\end{equation}
Figure \ref{subfig:HomogeneousSolution_SigmaCR_Polar_SM} shows the polar plot of $\sigma^{cr}$.
As indicated by \eqref{eq:CriticalStressStandardModel_simplified}, a modification of the $\hat{G}_{c}/\ell$ ratio leads once again to a homothetic transformation of the polar plot, whereas its shape is only controlled by the ratios of the anisotropic elastic constants. Therefore, with the SM the directional dependence of the critical stress cannot be flexibly calibrated.  

\subsection{Multi-damage model}

As follows, we derive and illustrate the strength surface and the directional critical stress of the MDM summarized in Section \ref{subsct:MultiDamageMechanismsModel}.

\subsubsection{Strength surfaces of the MDM}
\label{subsubsct:GoverningEqTwoMechanisms}
For the MDM, the damage energy release rate $G_i(\Bve, \underline{d})$ and the local damage resistance $R_i(d_i,\mathbf{0})$ associated with the damage mechanism $i=1,\dots,m$ take the form
\begin{equation}
G_i(\Bve, \underline{d}) = -\frac{1}{2}\, \boldsymbol{\varepsilon} : \IC_{,d_i}(\underline{d}) : \boldsymbol{\varepsilon}
\qquad\mbox{and}\qquad
R_i(d_i,\mathbf{0}) = \frac{\hat{G}_{c_i}}{c_{w_i} \ell_i}
\end{equation}
so that the admissible strain and stress domains read
\begin{align}
\calR_i(\underline{d}) &=
\left\{ \Bve \in \mathrm.Sym : 
-\frac{1}{2}\,\Bve : \IC_{,d_i}(\underline{d}) : \Bve
\;\leq\; \dfrac{\hat{G}_{c_i}}{c_{w_i}\ell_i} \right\}, \quad i=1,\ldots,m,\\[6pt]
\calS_i(\underline d) &=
\left\{ \boldsymbol{\sigma} \in \mathrm{Sym} :
-\frac{1}{2}\,   \IS(\underline{d}) : \Bsigma : \IC_{,d_i}(\underline{d}) : \IS(\underline{d}) : \Bsigma \;\leq\; \dfrac{\hat{G}_{c_i}}{c_{w_i}\ell_i} \right\},\quad i=1,\ldots,m,
\end{align}
with the damaged compliance tensor $\IS(d) = \IC^{-1}(d)$. We thus obtain a multi-surface formulation, with multiple admissible stress domains bounded by multiple strength surfaces given by
\begin{equation}
\partial \calS_i(\underline{0}) =
\left\{ \boldsymbol{\sigma} \in \mathrm{Sym} :
-\frac{1}{2}\,   \IS(\underline{0}) : \boldsymbol{\sigma} : \IC_{,d_i}(\underline{0}) : \IS(\underline{0}) : \boldsymbol{\sigma} \;   =\; \dfrac{\hat{G}_{c_i}}{c_{w_i}\ell_i}\right\},\quad i=1,\ldots,m.
\label{eq:DamageNucleationSurface_MDM}
\end{equation}

For the orthotropic 2D case with $m=2$, using Voigt notation, the derivatives of the degraded stiffness with respect to the two damage variables take the form
\begin{equation}
\begin{array}{lclc}
\underline{\IC}^{\scriptstyle{2D}}_{,d_1}(\underline{d}) = -
\begin{bmatrix}
2(1-d_1)\,C_{11} & (1-d_2)\,C_{12} & 0 \\
\hphantom{2}(1-d_2)\,C_{12} & 0 & 0     \\
0      & 0      & (1-d_2)\,C_{66}\\
\end{bmatrix}
\end{array}
\label{eq:DegradatedStiffnessMultipleDamage_1}
\end{equation}
and
\begin{equation}
\begin{array}{lclc}
\underline{\IC}^{\scriptstyle{2D}}_{,d_2}(\underline{d}) = -
\begin{bmatrix}
 0 & \hphantom{2}(1-d_1)\,C_{12} & 0     \\
(1-d_1)\,C_{12} & 2(1-d_2)\,C_{22} & 0     \\
0      & 0      & (1-d_1)\,C_{66}\\
\end{bmatrix}
\end{array}
\label{eq:DegradatedStiffnessMultipleDamage_2}
\end{equation}
and the admissible strain and stress domains read
\begin{align}
\begin{split}
(1-d_1) C_{11}\varepsilon_{11}^2 
+ (1-d_2) C_{12}\varepsilon_{11}\varepsilon_{22}
+ 2(1-d_2) C_{66}\varepsilon_{12}^2
\;&\leq\; \frac{\hat{G}_{c_1}}{c_{w_1}\ell_1},\\[6pt]
(1-d_1)C_{12}\varepsilon_{11}\varepsilon_{22}
+ (1-d_2)C_{22}\varepsilon_{22}^2
+ 2(1-d_1) C_{66}\varepsilon_{12}^2
\;&\leq\; \frac{\hat{G}_{c_2}}{c_{w_2}\ell_2},
\end{split}
\label{eq:ElasticDomains_StrainSpace_MDM}
\end{align}
\begin{align}
\begin{split}
\frac{S_{11}\sigma_{11}^2}{(1-d_1)^3}
+ \frac{S_{12}\sigma_{11}\sigma_{22}}{(1-d_1)^2(1-d_2)}
+ \frac{S_{66}\sigma_{12}^2}{2(1-d_1)^2(1-d_2)}
\;&\leq\; \frac{\hat{G}_{c_1}}{c_{w_1}\ell_1},\\[6pt]
\frac{S_{12}\sigma_{11}\sigma_{22}}{(1-d_1)(1-d_2)^2}
+ \frac{S_{22}\sigma_{22}^2}{(1-d_2)^3}
+ \frac{S_{66}\sigma_{12}^2}{2(1-d_1)(1-d_2)^2}
\;&\leq\; \frac{\hat{G}_{c_2}}{c_{w_2}\ell_2}.
\end{split}
\label{eq:ElasticDomains_StressSpace_MDM}
\end{align}
From these expressions it is straightforward to verify that strain hardening and stress softening are both satisfied for a fixed value of each damage variable with respect to the other one. 
Finally, the two strength surfaces take the form
\begin{align}
\begin{split}
S_{11}\sigma_{11}^2 + S_{12}\sigma_{11}\sigma_{22}
+ \frac{1}{2}S_{66}\sigma_{12}^2
= \frac{\hat{G}_{c_1}}{c_{w_1}\ell_1},
\\[6pt]
S_{12}\sigma_{11}\sigma_{22} + S_{22}\sigma_{22}^2
+ \frac{1}{2}S_{66}\sigma_{12}^2
= \frac{\hat{G}_{c_2}}{c_{w_2}\ell_2},
\end{split}
\label{eq:NucleationDomain_MDM}
\end{align}
from which we easily obtain the values of the critical stress components 
\begin{flalign}
\begin{array}{@{}l @{\qquad} l@{}}
\sigma_{11}^{\scriptstyle{cr}} =\sqrt{\dfrac{\hat{G}_{c_1}}{c_{w_1}\,\ell_1\,S_{11}}} &\quad\mbox{if}\quad \sigma_{22}=\sigma_{12}=0, \\[4mm]
\sigma_{22}^{\scriptstyle{cr}}=\sqrt{\dfrac{\hat{G}_{c_2}}{c_{w_2}\,\ell_2\,S_{22}}} &\quad\mbox{if}\quad \sigma_{11}=\sigma_{12}=0, \\[4mm]
\sigma_{12}^{\scriptstyle{cr}}=\min \left(\sqrt{\dfrac{2\hat{G}_{c_1}}{c_{w_1}\,\ell_1\,S_{66}}}, \sqrt{\dfrac{2\hat{G}_{c_2}}{c_{w_2}\,\ell_2\,S_{66}}} \,\right) &\quad\mbox{if}\quad \sigma_{11}=\sigma_{22}=0.
\end{array}
\label{eq:sigma_cr_axis_MDM}
\end{flalign}
In the original paper proposing this model~\cite{bleyer+alessi18}, the two damage variables $d_1$ and $d_2$ are interpreted as being respectively associated to longitudinal and transverse fracture mechanisms e.g. in fiber-reinforced composites, so that the two equations \eqref{eq:NucleationDomain_MDM}  represent the corresponding strength surfaces; they are illustrated in Figures~\ref{subfig:StrengthSurface_MDM_1} and~\ref{subfig:StrengthSurface_MDM_2} in the $\sigma_{11}-\sigma_{22}$ plane assuming $\sigma_{12}=0$. A modification of the $\hat{G}_{c_1}/\ell_1$ or of the $\hat{G}_{c_2}/\ell_2$ ratio leads to a non-homothetic transformation of the respective strength surface while leaving the other one unchanged. Flexibility in strength calibration under multiaxial stress states is still limited within the range of a single mechanism, however it is enhanced by the interplay of the two mechanisms which can be separately controlled. Note that the overall strength surface exhibits kinks at the transition points between the surfaces of the different mechanisms. This may be representative of the multiaxial strength response of some materials, whereas for others it may be more realistic to expect a gradual transition. Eq. \eqref{eq:sigma_cr_axis_MDM} shows that the ratio of critical stress components controlled by the same damage mechanism (e.g. $\sigma_{11}^{cr}/\sigma_{12}^{cr}$ if $\hat{G}_{c_1}/\ell_1<\hat{G}_{c_2}/\ell_2$) is still only dependent on the ratio of the relevant elastic compliance values, however different critical stress components are or may be associated to different damage mechanisms (e.g. $\sigma_{11}^{cr}$  and $\sigma_{12}^{cr}$), which provides additional flexibility.

\subsubsection{Directional critical stress of the MDM}
\begin{figure}[h!]
  \centering
  \begin{minipage}{\textwidth}
  \begin{subfigure}{0.45\textwidth}
      \centering
      \includegraphics[width=\textwidth]{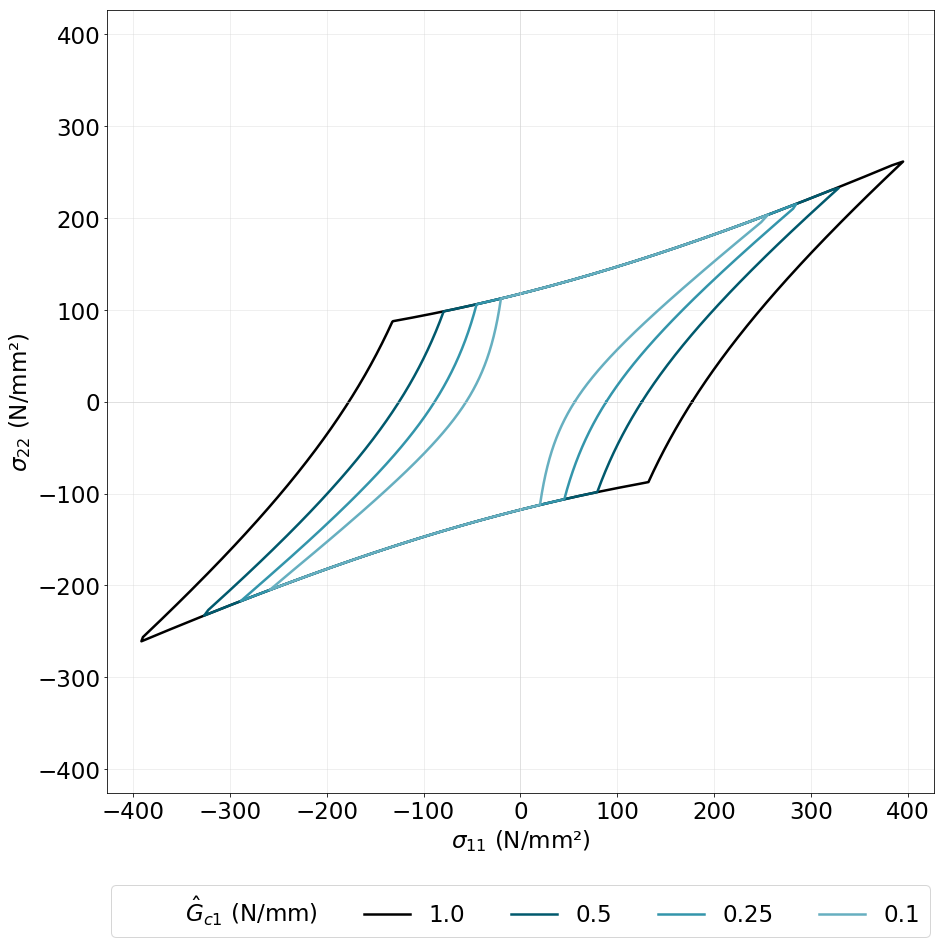}
      \caption{Strength surfaces of the MDM for varying $\hat{G}_{c_1}$ at fixed $\hat{G}_{c_2}\!\!=\!1.0$ N/mm (assuming $\sigma_{12}=0$).}
      \label{subfig:StrengthSurface_MDM_1}
    \end{subfigure}
    \hfill
    \begin{subfigure}{0.45\textwidth}
      \centering
      \includegraphics[width=\textwidth]{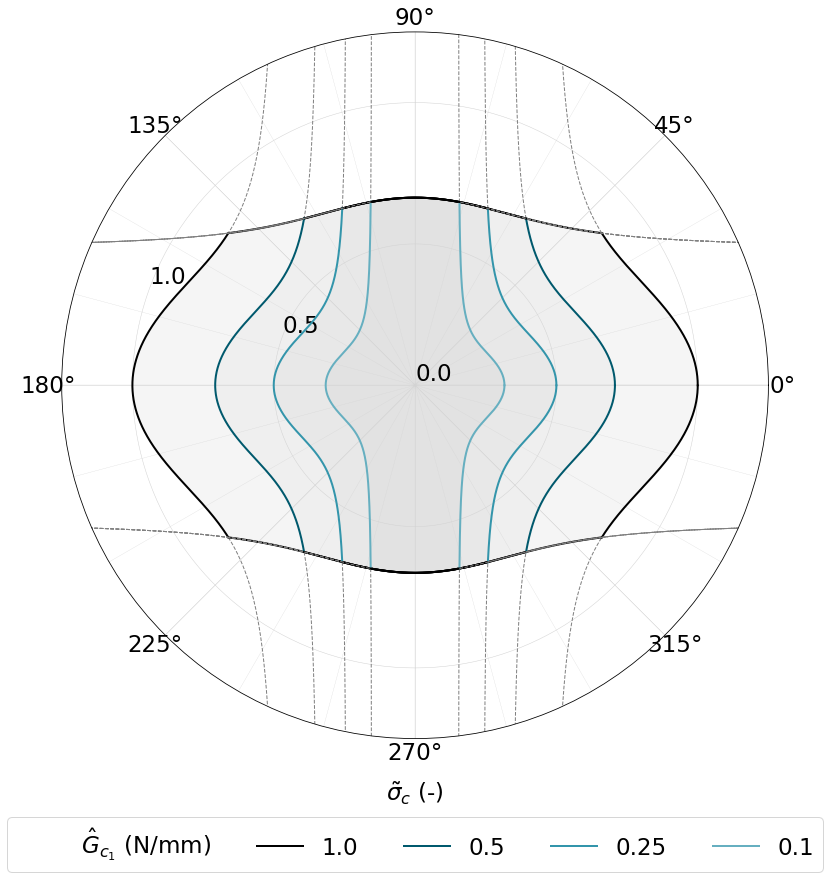}
      \caption{Polar plot of the normalized directional critical stress $\tilde{\sigma}^{\scriptstyle{cr}}(\theta) \!=\! \sigma^{\scriptstyle{cr}}(\theta) / \sigma^{\scriptstyle{cr}}(\theta \!=\! 0^\circ, \hat{G}_{c_1} \!\!=\! 1.0 \, \mbox{N/mm})$ for varying $\hat{G}_{c_1}$ at fixed $\hat{G}_{c_2}\!\!=\!1.0$ N/mm.}
      \label{subfig:HomogeneousSolution_SigmaCR_Polar_BA-Gc1}
    \end{subfigure}
  \end{minipage}
  \begin{minipage}{\textwidth}
    \begin{subfigure}{0.45\textwidth}
      \centering
     \includegraphics[width=\textwidth]{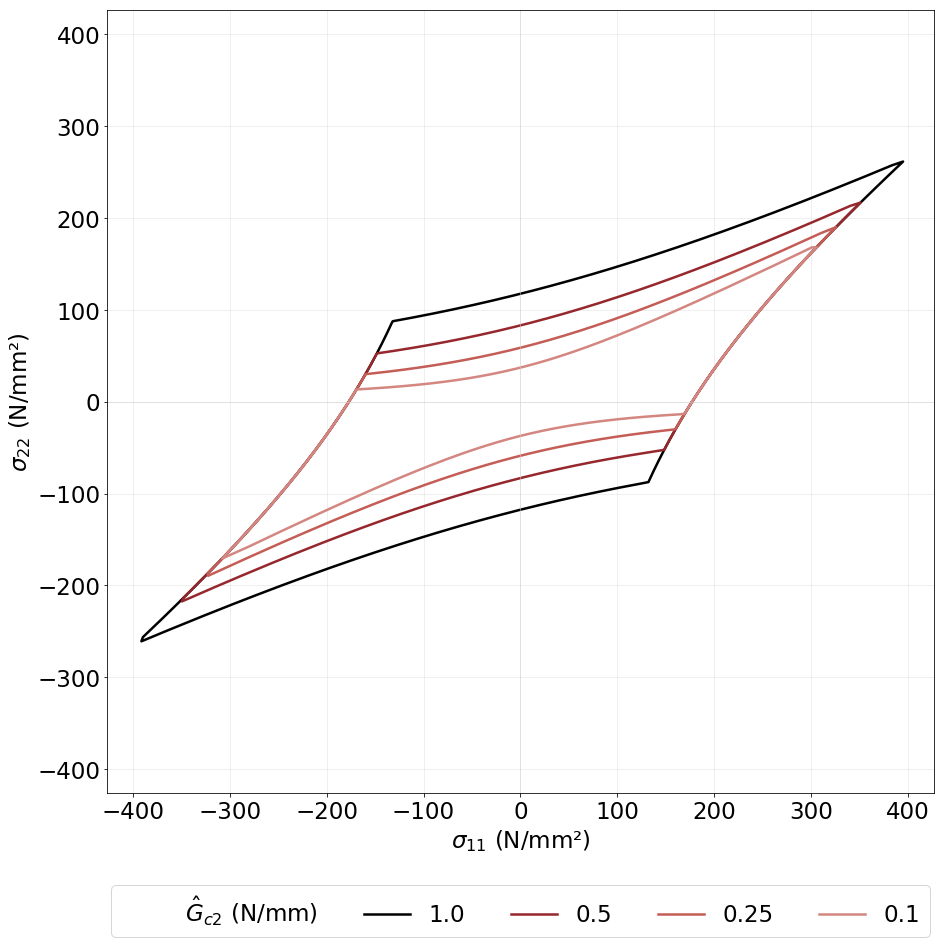}
      \caption{Strength surfaces of the MDM for varying $\hat{G}_{c_2}$ at fixed $\hat{G}_{c_1}\!\!=\!1.0$ N/mm (assuming $\sigma_{12}=0$).}
      \label{subfig:StrengthSurface_MDM_2}
    \end{subfigure}
    \hfill
    \begin{subfigure}{0.45\textwidth}
      \centering
      \includegraphics[width=\textwidth]{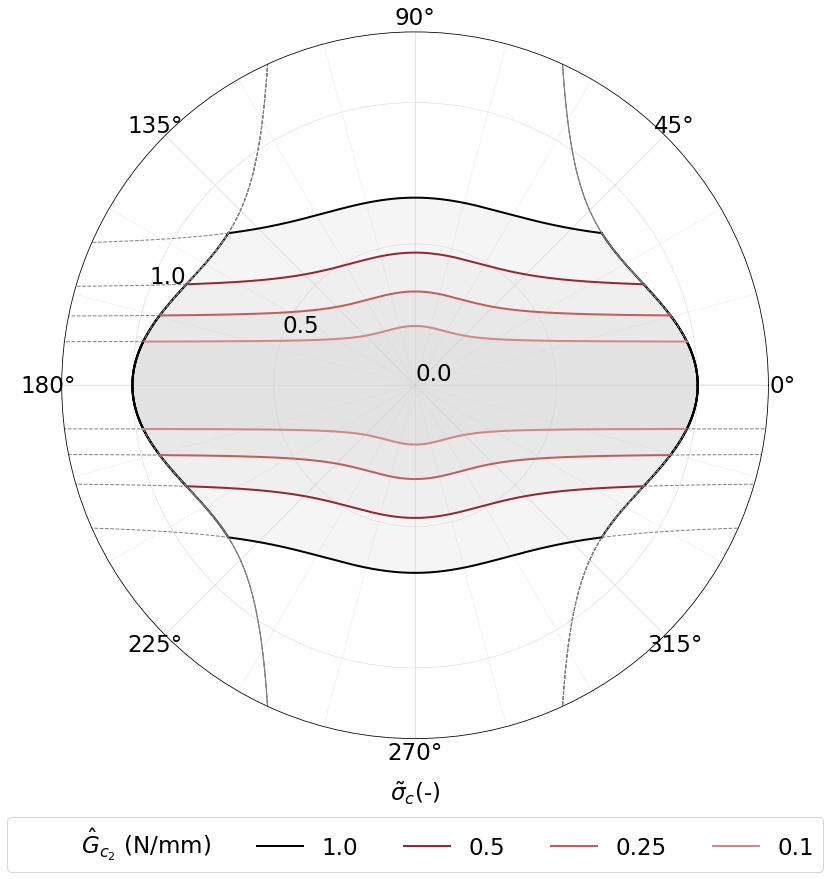}
      \caption{Polar plot of the normalized directional critical stress $\tilde{\sigma}^{\scriptstyle{cr}}(\theta) \!=\! \sigma^{\scriptstyle{cr}}(\theta) / \sigma^{\scriptstyle{cr}}(\theta \!=\! 0^\circ, \hat{G}_{c_2} \!\!=\! 1.0 \, \mbox{N/mm})$ for varying $\hat{G}_{c_2}$ at fixed $\hat{G}_{c_1}\!\!=\!1.0$ N/mm.}
      \label{subfig:HomogeneousSolution_SigmaCR_Polar_BA-Gc2}
    \end{subfigure}
  \end{minipage}
  \caption{Strength surface and polar plot of the directional critical stress according to the \textit{MDM}.}
\label{Figure:HomogeneousSolution_BA}
\end{figure}

The strength surfaces (\ref{eq:NucleationDomain_MDM}), upon insertion of the rotated uniaxial stress state \eqref{1D_rotated},  deliver the following non-smooth representation of the directional critical stress 
\begin{equation}  
\sigma^{\scriptstyle{cr}}(\theta) = \min \left(\sqrt{\dfrac{\hat{G}_{c_1}}{c_{w_1} \ell_1 S_1(\theta)}}, \sqrt{\dfrac{\hat{G}_{c_2}}{c_{w_2} \ell_2 S_2(\theta)}} \,\right),
\label{eq:CriticalStressMDM}
\end{equation}
where
\begin{equation}
    S_1(\theta) = 
    S_{11}\cos^4\theta + \left(S_{12} + \frac{S_{66}}{2}\right)\sin^2\theta\cos^2\theta,
\label{eq:S1}
\end{equation}
\begin{equation}
    S_2(\theta) = 
    S_{22}\sin^4\theta + \left(S_{12} + \frac{S_{66}}{2}\right)\sin^2\theta\cos^2\theta.
\label{eq:S1}
\end{equation}
Figures~\ref{subfig:HomogeneousSolution_SigmaCR_Polar_BA-Gc1} and ~\ref{subfig:HomogeneousSolution_SigmaCR_Polar_BA-Gc2} show the polar plots of $\sigma^{cr}$, respectively 
for a constant $\hat{G}_{c_2}/\ell_2$ with varying $\hat{G}_{c_1}/\ell_1$ and for a constant  $\hat{G}_{c_1}/\ell_1$ with varying $\hat{G}_{c_2}/\ell_2$. 
These plots lead to similar conclusions as the strength surfaces: the MDM allows for more flexibility than the SM in calibrating the orientation-dependent critical stress due to two independent scaling factors  corresponding to the two different mechanisms.
For a given loading angle, the directional critical stress is affected either only by $\hat{G}_{c_1}/\ell_1$ or only by $\hat{G}_{c_2}/\ell_2$, depending on the controlling mechanism, and the transition between ranges of angles affected by the two mechanisms is non-smooth.
Moreover, within the range of a given mechanism $i$, the quantitative effect of a change in $\hat{G}_{c_i}/\ell_i$ does not follow a homothetic scaling. 

\subsection{Proposed multi-cohesive model}
\label{subsubsct:GoverningEqNewModel}

As follows, we derive and illustrate the strength surface and the directional critical stress of our MCM presented in Section \ref{subsct:ProposedModel}.

\subsubsection{Strength surface of the proposed MCM}
For the MCM, the damage energy release rate $G(\Bve,d)$ and the local damage resistance $R(d,\mathbf{0})$ read
\begin{equation}
G(\Bve, d) = -\frac{1}{2}\, \Bve : \IC_{,d}(d) : \Bve
\qquad\mbox{and}\qquad
R(d,\mathbf{0}) = \frac{\hat{G}_{c}}{c_{w} \ell}
\end{equation}
as in the SM, 
and the admissible strain and stress domains are given by
\begin{align}
\calR(d) &= \left\{ \boldsymbol{\varepsilon} \in \mathrm{Sym} \; : \;
-\frac{1}{2}\,\boldsymbol{\varepsilon} : \IC_{,d}(d) : \boldsymbol{\varepsilon}
\leq \frac{\hat{G}_c}{c_w\ell} \right\}, \\[6pt]
\calS(d) &= \left\{ \boldsymbol{\sigma} \in \mathrm{Sym} : 
-\frac{1}{2}\, \IS(d) : \boldsymbol{\sigma} : \IC_{,d}(d) : \IS(d) : \boldsymbol{\sigma}
\leq \frac{\hat{G}_c}{c_w\ell} \right\},
\label{eq:StrengthSurface_MCM}
\end{align}
leading to the strength surface
\begin{equation}
\partial \calS(0)
= \left\{ \Bsigma \in \mathrm{Sym} :
\; -\frac{1}{2}\IS(0) : \Bsigma : \IC_{,d}(0): \IS(0) : \boldsymbol{\sigma} = \frac{\hat{G}_c}{c_w \ell} \right\}.
\label{eq:DamageNucleationSurface_MCM}
\end{equation}

\begin{figure}[htbp]
  \centering
  \begin{minipage}[t]{0.45\textwidth}
    \begin{subfigure}{\textwidth}
      \centering
      \includegraphics[scale=0.25]{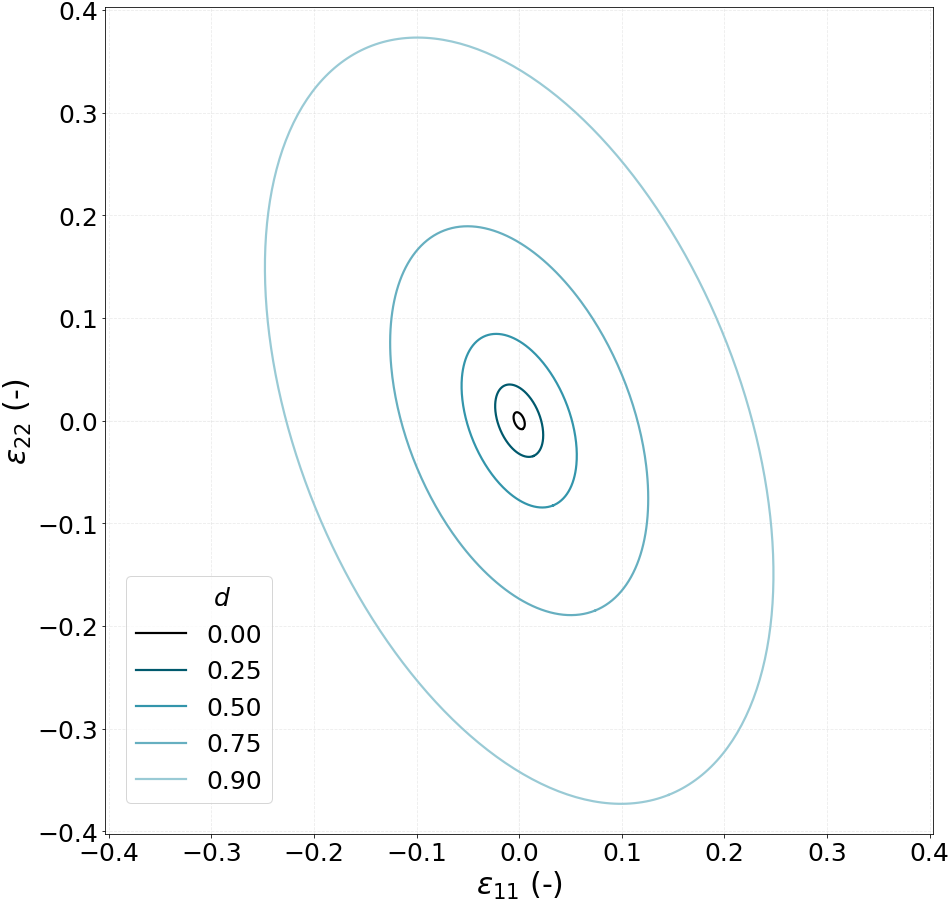}
      \
      \caption{Admissible strain domains $\calR(d)$ (assuming $\varepsilon_{12}=0$).}
      \label{subfig:ElasticDomain_StrainSpace_MCM}
    \end{subfigure}
  \end{minipage}
  \hfill
  \begin{minipage}[t]{0.45\textwidth}
    \begin{subfigure}{\textwidth}
     \centering
     \includegraphics[scale=0.25]{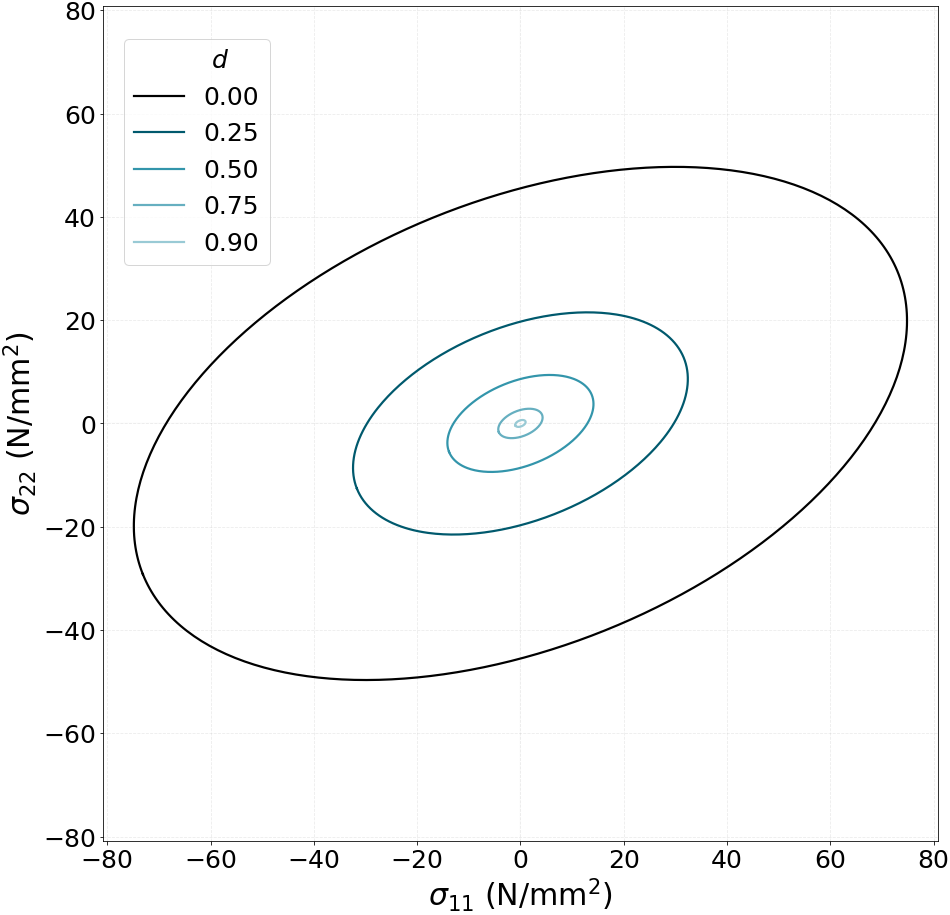}
      \caption{Admissible stress domains $\calS(d)$ (assuming $\sigma_{12}=0$).}
      \label{subfig:ElasticDomain_StressSpace_MCM}
    \end{subfigure}
  \end{minipage}
  \caption{MCM admissible strain and stress domains for the orthotropic 2D case.}
  \label{Figure:ElasticDomains_MCM}
\end{figure}

In the orthotropic 2D setting in Voigt notation, the derivative of the degraded stiffness with respect to the damage variable reads
\begin{equation}
\begin{array}{lcl}
\underline{\IC}^{\scriptstyle{2D}}_{,d}(d) =
\begin{bmatrix}
g_1'(d)\,C_{11} &  g_{12}'(d)\,C_{12} & 0  \\
g_{12}'(d)\,C_{12} &  g_2'(d)\,C_{22} & 0  \\
0  &  0  &  g_{3}'(d)\,C_{66} \\
\end{bmatrix}
\end{array}
\label{eq:StiffnessDerivativeNewModel}
\end{equation}
with $g_{3}(d) = \sqrt{g_1(d)\,g_2(d)}$, and 
the admissible strain and stress domains are obtained as follows
\begin{align}
\begin{split}
f(d) \left[ \frac{C_{11}r_1}{f_1^2(d)}\varepsilon_{11}^2 + \frac{2C_{12}r_{m} f_{12}(d)}{f_1^{3/2}(d)f_2^{3/2}(d)}\varepsilon_{11}\varepsilon_{22}+ \frac{C_{22}r_2}{f_2^2(d)}\varepsilon_{22}^2 + \frac{4C_{66}r_{m} f_{12}(d)}{f_1^{3/2}(d)f_2^{3/2}(d)}\varepsilon_{12}^2 \right]
\;&\leq\; \frac{\hat{G}_{c}}{c_{w}\ell},
\end{split}
\end{align}
\begin{align}
\begin{split}
\frac{f(d)}{(1-d)^4} \left[ \frac{S_{11}}{C_{11}C_{22}-C_{12}^2} \right. & \left( C_{11}C_{22}r_1 - 2C_{12}^2r_{m}\frac{f_{12}(d)}{f_2(d)} + C_{12}^2r_2\frac{f_{1}(d)}{f_2(d)} \right)\sigma_{11}^2 \\[4mm]
+ \frac{S_{22}}{C_{11}C_{22}-C_{12}^2} & \left( C_{11}C_{22}r_2 - 2C_{12}^2r_{m}\frac{f_{12}(d)}{f_1(d)} + C_{12}^2r_1\frac{f_{2}(d)}{f_1(d)} \right) \sigma_{22}^2 +\frac{S_{66}f_{12}(d)}{\sqrt{f_1(d)f_2(d)}}r_{m}\sigma_{12}^2 \\[4mm]
+ \frac{2S_{12}\sqrt{f_1(d)f_2(d)}}{C_{11}C_{22}-C_{12}^2} & \left.  \left( C_{11}C_{22}\frac{r_1}{f_1(d)} + C_{11}C_{22}\frac{r_2}{f_2(d)} - (C_{11}C_{22}+C_{12}^2)\frac{r_{m} f_{12}(d)}{f_1(d)f_2(d)} \right) \sigma_{11}\sigma_{22} \right] \;\leq\; \frac{\hat{G}_{c}}{c_{w}\ell},
\end{split}
\end{align}
with the abbreviations
\begin{flalign}
\begin{array}{@{}l @{\qquad} l@{}}
f(d) = (1-d)(1+d+2pd), & \\[4mm]
f_i(d) = (1-d)^2 + 2r_i d (1+pd) & \mbox{with}\qquad r_i = \dfrac{\ell_{c_i}}{\ell},\qquad i=1,2, \\[4mm]
f_{12}(d) = (1-d)^2 + \dfrac{2r_1r_2}{r_{m}} d (1+pd) & \mbox{with}\qquad r_{m} = \dfrac{1}{2}(r_1+r_2).
\end{array}
\label{eq:abbreviationsMCM}
\end{flalign}
Exemplary admissible strain and stress domains are depicted in Figure~\ref{Figure:ElasticDomains_MCM} and display respectively strain hardening and stress softening behavior for all values of $d\in[0,1)$.
The strength surface takes the form
\begin{equation}
\label{strength_MCM}
S_{11}r_1\sigma_{11}^2+ S_{22}r_2\sigma_{22}^2 + 2S_{12}r_{m}\sigma_{11}\sigma_{22} + S_{66}r_{m}\sigma_{12}^2 = \frac{\hat{G}_c}{c_w \ell}
\end{equation}
from which we obtain the values of the critical stress components as follows
\begin{flalign}
\begin{array}{@{}l @{\qquad} l@{}}
\sigma_{11}^{\scriptstyle{cr}}=\sqrt{\dfrac{\hat{G}_c}{c_w\,\ell_{c_1}\,S_{11}}} &\quad\mbox{if}\quad \sigma_{22}=\sigma_{12}=0, \\[4mm]
\sigma_{22}^{\scriptstyle{cr}}=\sqrt{\dfrac{\hat{G}_c}{c_w\,\ell_{c_2}\,S_{22}}} &\quad\mbox{if}\quad \sigma_{11}=\sigma_{12}=0, \\[4mm]
\sigma_{12}^{\scriptstyle{cr}} =\sqrt{\dfrac{\hat{G}_c}{c_w\,\ell_{c_m}\,S_{66}}} &\quad\mbox{if}\quad \sigma_{11}=\sigma_{22}=0.
\end{array}
\label{eq:sigma_cr_axis_MCM}
\end{flalign}
The strength surface \eqref{strength_MCM} is illustrated in Figures~\ref{subfig:StrengthSurface_MCM_1} and~\ref{subfig:StrengthSurface_MCM_2} where we respectively modify $r_1$ at fixed $r_2$ or $r_2$ at fixed $r_1$. With the MCM, flexibility in strength calibration under multiaxial stress states is introduced by the additional variables $r_1$ and $r_2$ (or, equivalently for a fixed $\ell$, by the cohesive lengths $l_{c_1}$ and $l_{c_2}$), which allow for a change in shape of the strength surface. Note that we again have only one smooth strength surface as with the SM. Eq. \eqref{eq:sigma_cr_axis_MCM} shows that the ratio of the critical stress components depends not only on the ratio of the relevant elastic compliance values, but also on the ratio of the relevant cohesive lengths, which provides additional flexibility with respect to the SM.

\subsubsection{Directional critical stress of the proposed MCM}
\label{subsct:HomogeneousResponseNewModel}
\begin{figure}[h!]
  \centering
  \begin{minipage}{\textwidth}
    \begin{subfigure}{0.45\textwidth}
      \centering
        \includegraphics[width=\textwidth]{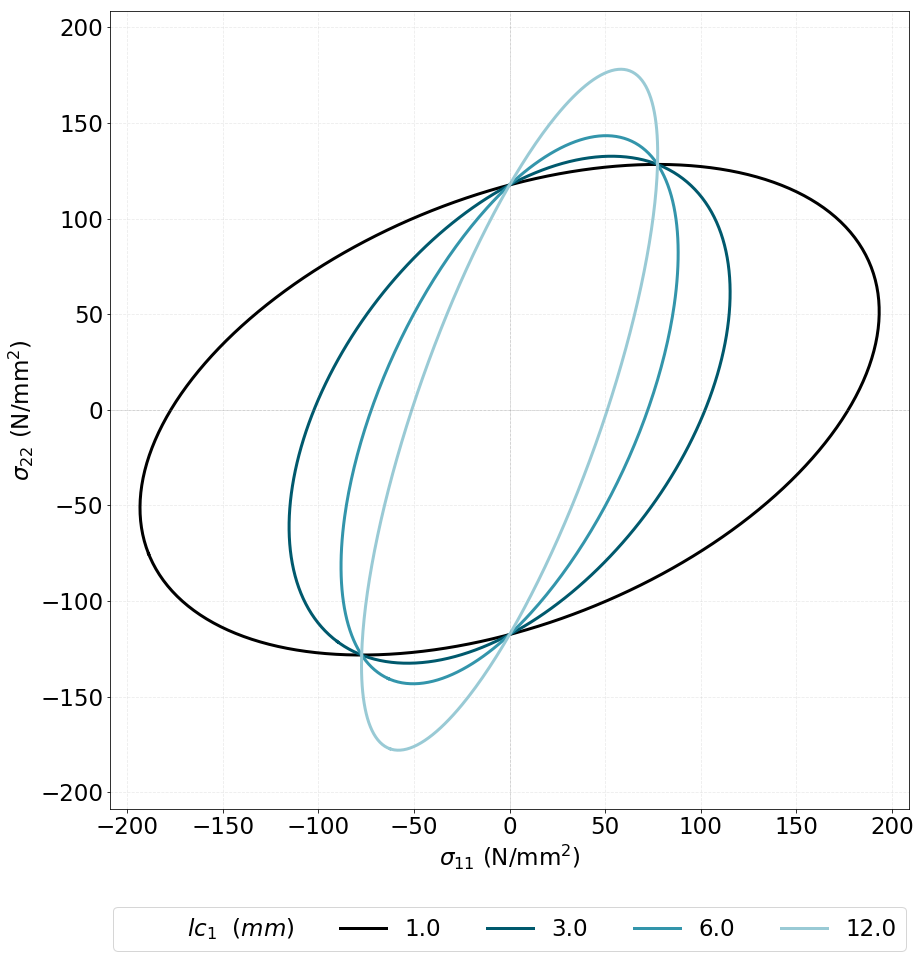}
      \caption{Strength surfaces of the MCM for varying cohesive length $l_{c_1}$ at fixed $\ell_{c_2}\!\!=\!1.0$ mm (assuming $\sigma_{12}=0$).}
      \label{subfig:StrengthSurface_MCM_1}
    \end{subfigure}
    \hfill
    \begin{subfigure}{0.45\textwidth}
      \centering
      \includegraphics[width=\textwidth]{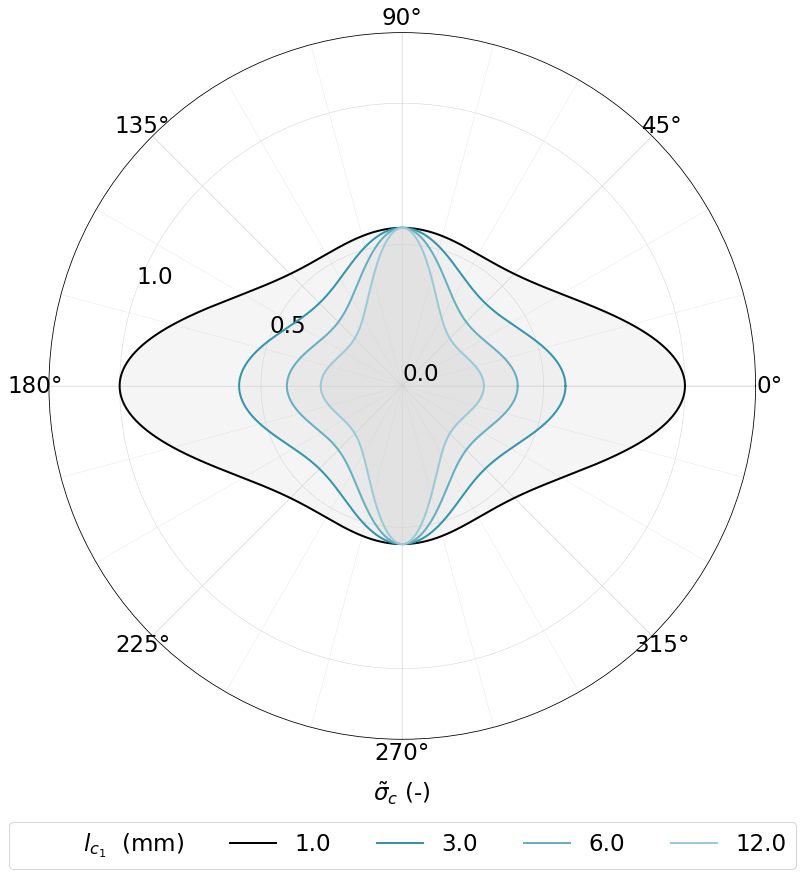}
      \caption{Polar plot of the normalized directional critical stress $\tilde{\sigma}^{\scriptstyle{cr}}(\theta) \!=\! \sigma^{\scriptstyle{cr}}(\theta) / \sigma^{\scriptstyle{cr}}(\theta \!=\! 0^\circ, \ell_{c_1} \!\!=\! 1.0 \, \mbox{mm})$ for varying cohesive length $\ell_{c_1}$ at fixed $\ell_{c_2}\!\!=\!1.0$ mm.}
      \label{subfig:HomogeneousSolution_SigmaCR_Polar_NM-Lc1}
    \end{subfigure}
  \end{minipage}
  \begin{minipage}{\textwidth}
    \begin{subfigure}{0.45\textwidth}
      \centering
      \includegraphics[width=\textwidth]{plots/StrengthSurfaces/MCM/StrengthSurface_MCM_1.png}    \caption{Strength surfaces of the MCM for varying cohesive length $l_{c_2}$ at fixed $\ell_{c_1}\!\!=\!1.0$~mm (assuming $\sigma_{12}=0$).}
      \label{subfig:StrengthSurface_MCM_2}
    \end{subfigure}
    \hfill
    \begin{subfigure}{0.45\textwidth}
      \centering
      \includegraphics[width=\textwidth]{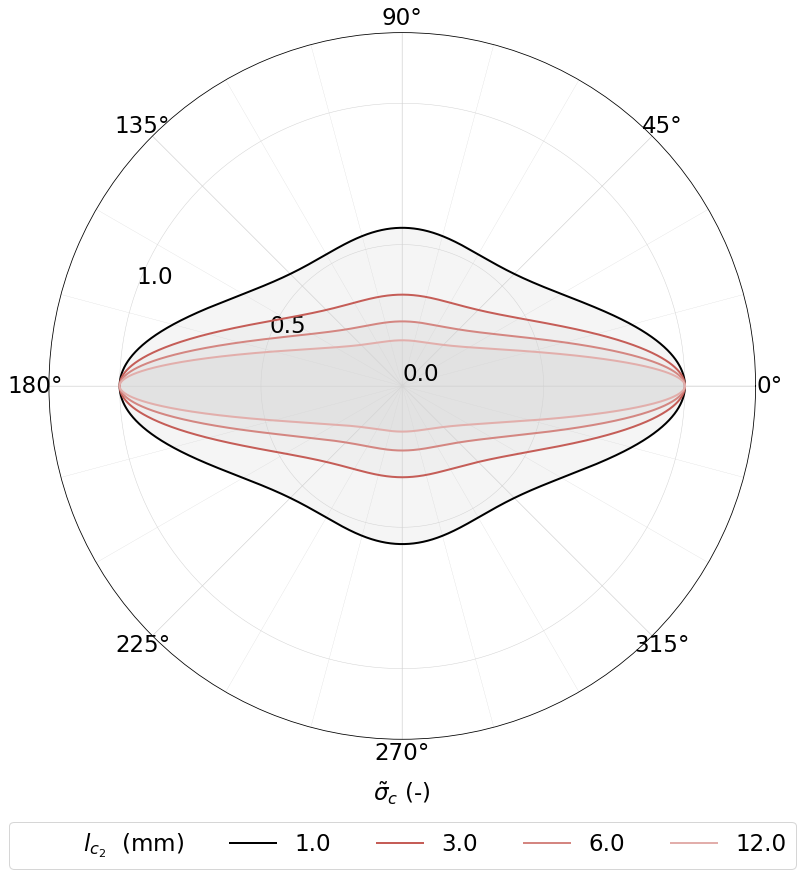}
      \caption{Polar plot of the normalized directional critical stress $\tilde{\sigma}^{\scriptstyle{cr}}(\theta) \!=\! \sigma^{\scriptstyle{cr}}(\theta) / \sigma^{\scriptstyle{cr}}(\theta \!=\! 0^\circ, \ell_{c_2} \!\!=\! 1.0 \, \mbox{mm})$ for varying cohesive length $\ell_{c_2}$ at fixed $\ell_{c_1}\!\!=\!1.0$~mm.}
      \label{subfig:HomogeneousSolution_SigmaCR_Polar_NM-Lc2}
    \end{subfigure}
  \end{minipage}
  \caption{Strength surface and polar plot of the directional critical stress according to the \textit{MCM}.}
  \label{Figure:HomogeneousSolution_NM}
\end{figure}

Insertion of the rotated uniaxial stress state \eqref{1D_rotated} in the strength surface ~\eqref{strength_MCM} yields the directional critical stress

\begin{equation}
    \sigma^{\scriptstyle{cr}}(\theta) = \sqrt{\displaystyle \frac{\hat{G}_c}{c_w \ell S(\theta)} }
\label{eq:CriticalStressThetaMCM}
\end{equation}
where
\begin{equation}
    S(\theta) =
    S_{11}r_1\cos^4\theta + S_{22}r_2\sin^4\theta + \left(2S_{12} + S_{66}\right)r_m\sin^2\theta\cos^2\theta.
\label{eq:CriticalStressMCM_simplified}
\end{equation}
Although \eqref{eq:CriticalStressMCM_simplified} looks very similar to \eqref{eq:CriticalStressSM_simplified}, due to the factors $r_1$, $r_2$, $r_m$ in \eqref{eq:CriticalStressMCM_simplified} 
the ratio of the cohesive lengths in different material directions affects the ratio of the critical stresses at different loading angles. Thus, more flexibility is added with respect to the SM, not due to different mechanisms as in the MDM but to different cohesive behaviors in different material directions. Figures~\ref{subfig:HomogeneousSolution_SigmaCR_Polar_NM-Lc1} and~\ref{subfig:HomogeneousSolution_SigmaCR_Polar_NM-Lc2} visualize the polar plot of $\sigma^{cr}$ under a change of $\ell_{c_1}$ with fixed $\ell_{c_2}$ and under a change of $\ell_{c_2}$ with fixed $\ell_{c_1}$, respectively. 
%

\subsection{Numerical verification of the strength surface for the proposed MCM}
\begin{figure}[h!]
 \centering
    \includegraphics[width=0.35\textwidth]{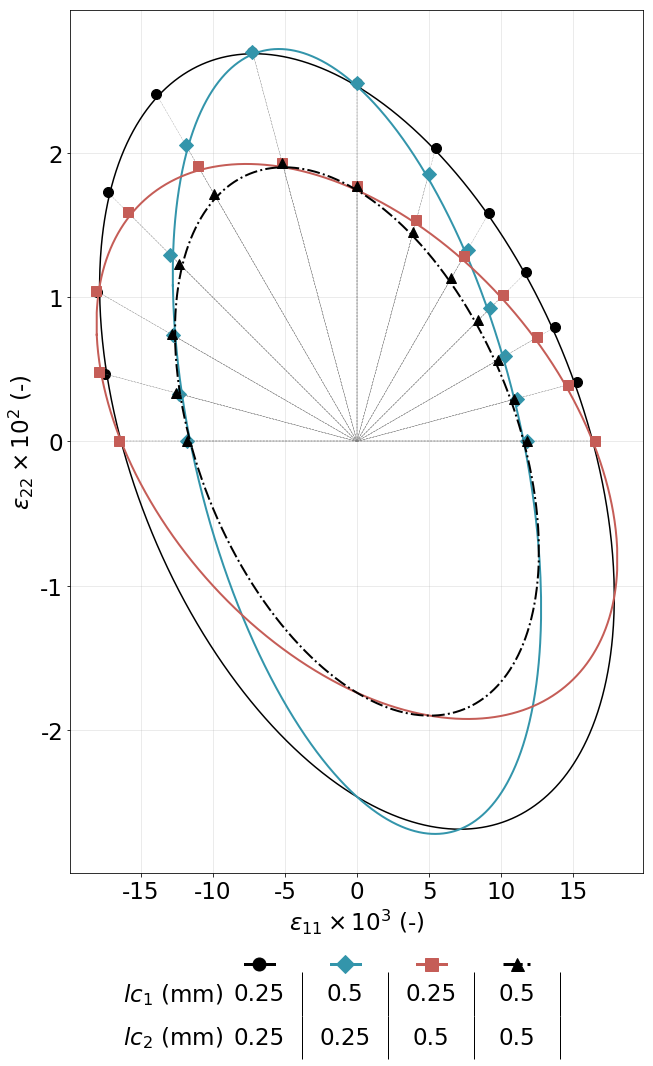}
      \caption{Numerical verification of~\eqref{strength_surface_strain} (assuming $\varepsilon_{12} =0$) for different combinations of the cohesive lengths $l_{c_1}$, $l_{c_2}$. Markers indicate numerical results for different $\Theta$.}
      \label{fig:NumericalVerification}
\end{figure}

As a verification, we now aim at numerically reproducing the theoretical strength surface. Since the analytical derivations of this paper are carried out in 2D while the numerical implementation is in plane-strain conditions, we reproduce the strength surface in the strain space, i.e. the boundary of the admissible strain domain for $d=0$ given by

\begin{equation}
\label{strength_surface_strain}
 C_{11}r_1\varepsilon_{11}^2 + C_{22}r_2\varepsilon_{22}^2 + 2C_{12}r_{m}\varepsilon_{11}\varepsilon_{22}+  4C_{66}r_{m}\varepsilon_{12}^2
= \frac{\hat{G}_{c}}{c_{w}\ell}.
\end{equation}
To this end, we follow the approach proposed by Vicentini et al.~\cite{vicentini2025variational} and use a square domain of edge length $L = 1$ mm subjected to Dirichlet boundary conditions  on opposite edges, with prescribed displacements $U_1 = U \cos(\Theta)$ in the $1$-direction and $U_2 = U \sin(\Theta)$ in the $2$-direction, where $\Theta$ is a fixed loading angle and $U$ is uniformly incremented. This loading case ensures initially homogeneous strain fields with components $\varepsilon_{11} = 2U_1/L$, $\varepsilon_{22} = 2U_2/L$, and $\varepsilon_{12} = 0$, such that the strain ratio is controlled by the angle $\Theta$ according to $\varepsilon_{22}/\varepsilon_{11} = \tan(\Theta)$. For each value of \(\Theta\), we numerically identify the time step at which the phase-field variable becomes non-zero for the first time (using the tolerance \(\max_{\boldsymbol{x} \in \Omega} d(\boldsymbol{x}) > 10^{-6}\)) and record the corresponding strain value. These values are visualized with discrete markers in Figure \ref{fig:NumericalVerification} along with the theoretical curves shown as continuous lines, for different values of the two cohesive lengths. We observe that the theoretical strength surface is well matched by the numerical results in all cases. 
%
%
\section{Crack propagation with the proposed multi-cohesive model}
\label{sct:LocalizedSolution}
In Section \ref{sct:3_HomogeneousState} we studied our proposed MCM, in comparison with the SM and the MDM, from the standpoint of crack nucleation. In this section, we focus on localization of the phase-field variable and crack propagation. As this setting is not amenable to closed-form solutions, we illustrate numerical examples in three dimensions. For the orthotropic case in Voigt notation we then have
\begin{equation}
\begin{array}{lclc}
\underline{\IC}_{0}^{\scriptstyle{3D}} =
\begin{bmatrix}
C_{11} & C_{12} & C_{13} & 0              & 0              & 0      \\
C_{12} & C_{22} & C_{23} & 0              & 0              & 0      \\
C_{13} & C_{23} & C_{33} & 0              & 0              & 0      \\
0              & 0              & 0              & C_{44} & 0              & 0      \\
0              & 0              & 0              & 0              & C_{55} & 0      \\
0              & 0              & 0              & 0              & 0              & C_{66} 
\end{bmatrix}
\end{array}.
\label{eq:DegradatedStiffnessStandardModel_3D}
\end{equation}
The degradation tensor, defined in~(\ref{eq:DegradationTensorProposedModel_2D}) for the 2D case, now reads
\begin{equation}
\underline{\ID}^{\scriptstyle{3D}}(d) =
\begin{bmatrix}
\sqrt{g_1(d)} &              0 &             0 &                         0 &                        0 & 0\\
0             &  \sqrt{g_2(d)} &             0 &                         0 &                        0 & 0\\
0             &              0 & \sqrt{g_3(d)} &                         0 &                        0 & 0\\
0             &              0 &            0  &  \sqrt[4]{g_1(d)\,g_2(d)} &                        0 & 0\\
0             &              0 &            0  &                         0 & \sqrt[4]{g_1(d)\,g_3(d)} & 0\\
0             &              0 &            0  &                         0 &                        0 & \sqrt[4]{g_2(d)\,g_3(d)}
\end{bmatrix}
\label{eq:DegradationTensorProposedModel_3D}
\end{equation}
with the degradation functions given by (\ref{eq:degradationLorentz}) and the cohesive lengths $\ell_{c_i}$, now with $i=1,2,3$, granting flexibility to modify the shape of the strength surface in all three material directions.
\begin{table}
  \caption{Common and test-specific MCM material parameters used for the numerical examples in Section~\ref{sct:LocalizedSolution}.}
  \label{tab:merged_params_localized_vertical}
  \centering
  \scriptsize 
  \renewcommand{\arraystretch}{1.0} 
  \setlength{\tabcolsep}{4pt}       
  \begin{tabularx}{\textwidth}{@{}p{0.25\textwidth}X X X@{}} 
    \toprule
    \multicolumn{4}{@{}l@{}}{\textbf{Common parameters}} \\
    \midrule
    Elastic coefficients (N/mm$^{2}$) &
    \multicolumn{3}{@{}X@{}}{ 
        $
        \setlength\arraycolsep{3pt}
        \begin{array}{lll}
        C_{11}=14945 & C_{12}=3970 & C_{44}=1295 \\
        C_{22}=6582  & C_{13}=3970 & C_{55}=1297 \\
        C_{33}=6586  & C_{23}=4180 & C_{66}=1241
        \end{array}
        $
    } \\
    Cohesive degradation parameter (–) & \multicolumn{3}{@{}l@{}}{$p = 2.0$} \\
    Fracture energy scaling factor (N/mm) & \multicolumn{3}{@{}l@{}}{$\hat{G}_c = 1.0$} \\
    Crack orientation intensity (–) & \multicolumn{3}{@{}l@{}}{
        $
        \begin{array}{lll}
        a_{1} = 1 & a_{2} = 0 & a_{3} = 0
        \end{array}
        $
    } \\
    \midrule
    \multicolumn{2}{@{}l@{}}{\textbf{Three-dimensional uniaxial tension test}} &
    \multicolumn{2}{@{}l@{}}{\textbf{Plane-strain compact tension test}} \\
    \cline{1-2} 
    \cline{3-4} 
    Regularization length (mm) & $\ell = 0.15$ & Regularization length (mm) & $\ell = 0.035$ \\
    Cohesive lengths (mm) & $\ell_{c_1} = [1.0,3.0],\; \ell_{c_2}=\ell_{c_3}=[1.0,3.0]$ & Cohesive lengths (mm) & $\ell_{c_i}=0.2,\; i=1,2,3$ \\
    Crack orientation intensity (–) & $\alpha = [0.0,1.0,10.0]$ & Crack orientation intensity (–) & $\alpha = 10.0$ \\
    \midrule
    \multicolumn{2}{@{}l@{}}{\textbf{Plane-strain single-edge notched tension test with layered structure}} &
    \multicolumn{2}{@{}l@{}}{\textbf{Three-dimensional single-edge notched tension test with laminate structure}} \\
    \cline{1-2} 
    \cline{3-4} 
    Regularization length (mm) & $\ell=0.2$ & Regularization length (mm) & $\ell=1.25$ \\
    Cohesive lengths (mm) & $\ell_{c_i}=1.1,\; i=1,2,3$ & Cohesive lengths (mm) & $\ell_{c_i}=6.6,\; i=1,2,3$ \\
    Crack orientation intensity (–) & $\alpha = 5.0$ & Crack orientation intensity (–) & $\alpha = 10.0$ \\
    \bottomrule
  \end{tabularx}
\end{table}
As introduced for the SM in Section~\ref{subsct:StandardAnisotropicModel}, control over the orientation of the crack is obtained by the structural tensor $\BA$. Here, we design this tensor in such a way that it does not affect the critical stress at crack nucleation and that it enforces the crack direction to be aligned with a desired unit vector $\Ba$. In the local material coordinate system $\bar{\Be}_i$ the structural tensor reads
\begin{equation}
    \BA = \frac{1}{1+\alpha/3}
    \begin{bmatrix}
 1+\alpha\, {a}_{1}^2 & 0 & 0 \\
 0 & 1+\alpha\, {a}_{2}^2 & 0 \\
 0 & 0 & 1+\alpha\, {a}_{3}^2  \\
    \end{bmatrix},
\label{eq:Amatrix}
\end{equation}
where $\alpha$ represents the intensity with which the crack direction is enforced to align with the prescribed direction $\Ba$. For $\alpha=0$, the structural tensor degenerates to the identity, hence there is no preferential crack orientation. All material parameters chosen for the following examples are summarized in Table~\ref{tab:merged_params_localized_vertical}.
For the sake of simplicity we restrict ourselves to a preferred crack orientation in the first local coordinate direction $\Ba = \bar{\Be}_1$. We implement the discretized formulation of the model in the commercial finite element package Abaqus making use of the temperature-phase-field analogy ~\cite{Navidtehrani2021b}.

\subsection{Three-dimensional uniaxial tension test}
\label{subsct:UniaxialTT-LocalizedSolution}
\begin{figure}[h!]
 \centering
  \includegraphics[width=0.85\textwidth]{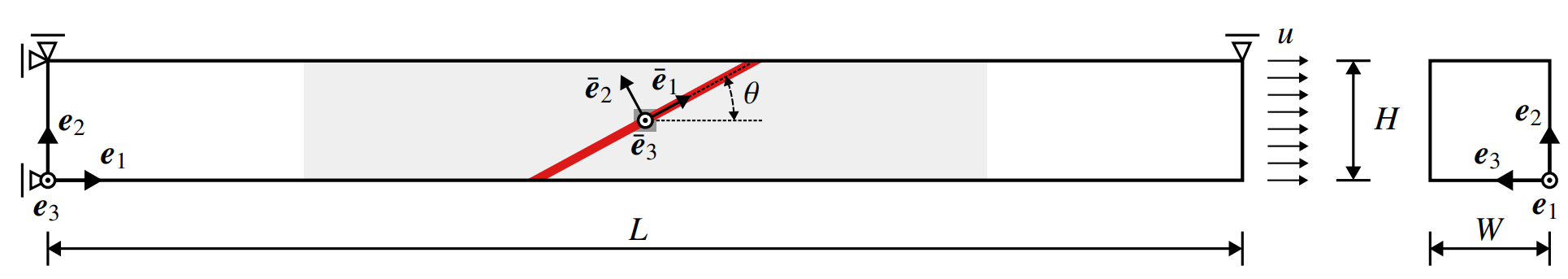}
   \caption{\textit{Three-dimensional uniaxial tension test.} The sample has dimensions $L = 10$~mm, $H = W = 1$~mm and is subjected to an imposed displacement $u = 0.2$~mm applied in increments $\Delta u = 1 \cdot 10^{-4}$ mm. Localization in the middle of the specimen is obtained by reducing  $\hat{G}_c$ by $5$\% in the dark gray element. The sample is meshed using 8-node brick elements with global seed size of $l_h = 0.03$~mm.}
   \label{Figure:Sketch_BarBVP}
 \end{figure}

In the first 3D example, we analyze the localization response of a bar subjected to a uniaxial tensile stress, see Figure~\ref{Figure:Sketch_BarBVP}. The length of the bar $L$ is sufficiently large (and the other parameters are also appropriately chosen) such that the homogeneous solution is expected to be unstable.  We enforce the crack to occur in the local direction $\bar{\Be}_1$ by setting $\mathbf{a}=\bar{\Be}_1$.
\begin{figure}
 \centering
  \begin{minipage}{\textwidth}
    \begin{subfigure}{0.45\textwidth}
      \centering
      \includegraphics[width=\textwidth]{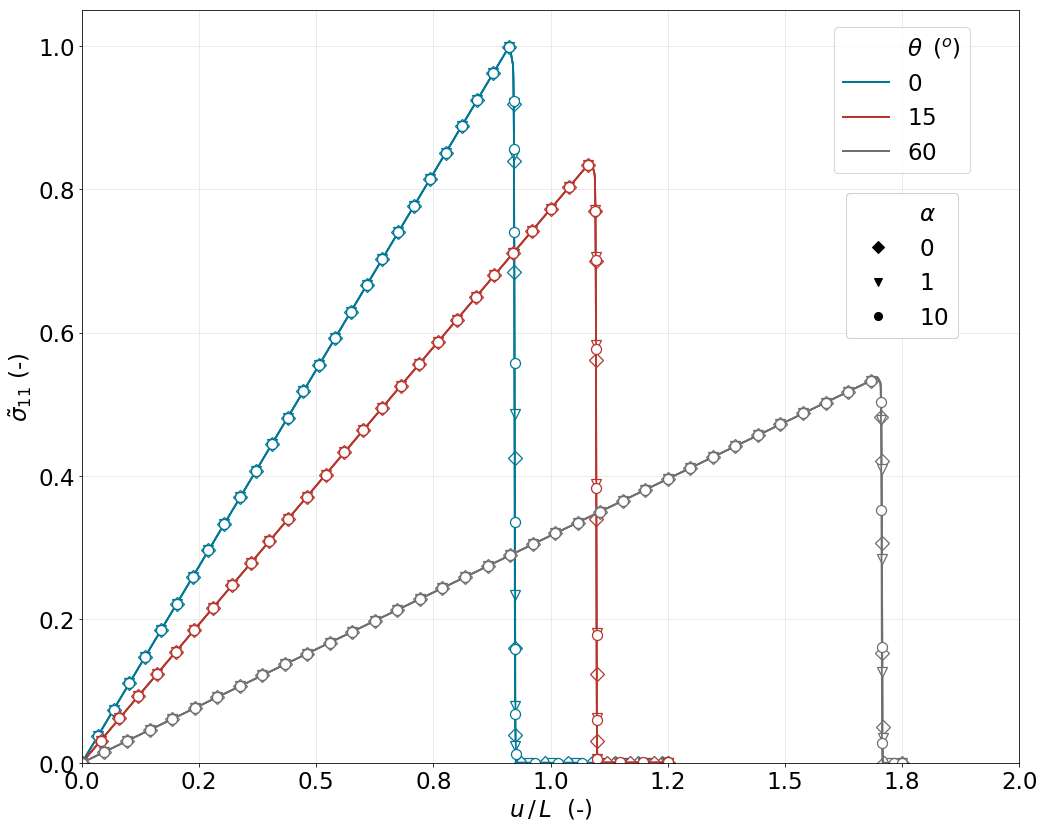}
      \caption{Normalized stress $\tilde{\sigma}_{11} = \sigma_{11}/\sigma_{cr}(\theta = 0^\circ,\alpha = 0)$ versus imposed displacement for different material orientations $\theta$ and values of $\alpha$.}
      \label{subfig:Stress_Strain_UDBar_alpha}
    \end{subfigure}
    \hfill
    \begin{subfigure}{0.49\textwidth}
      \centering
      \includegraphics[width=1.0\textwidth]{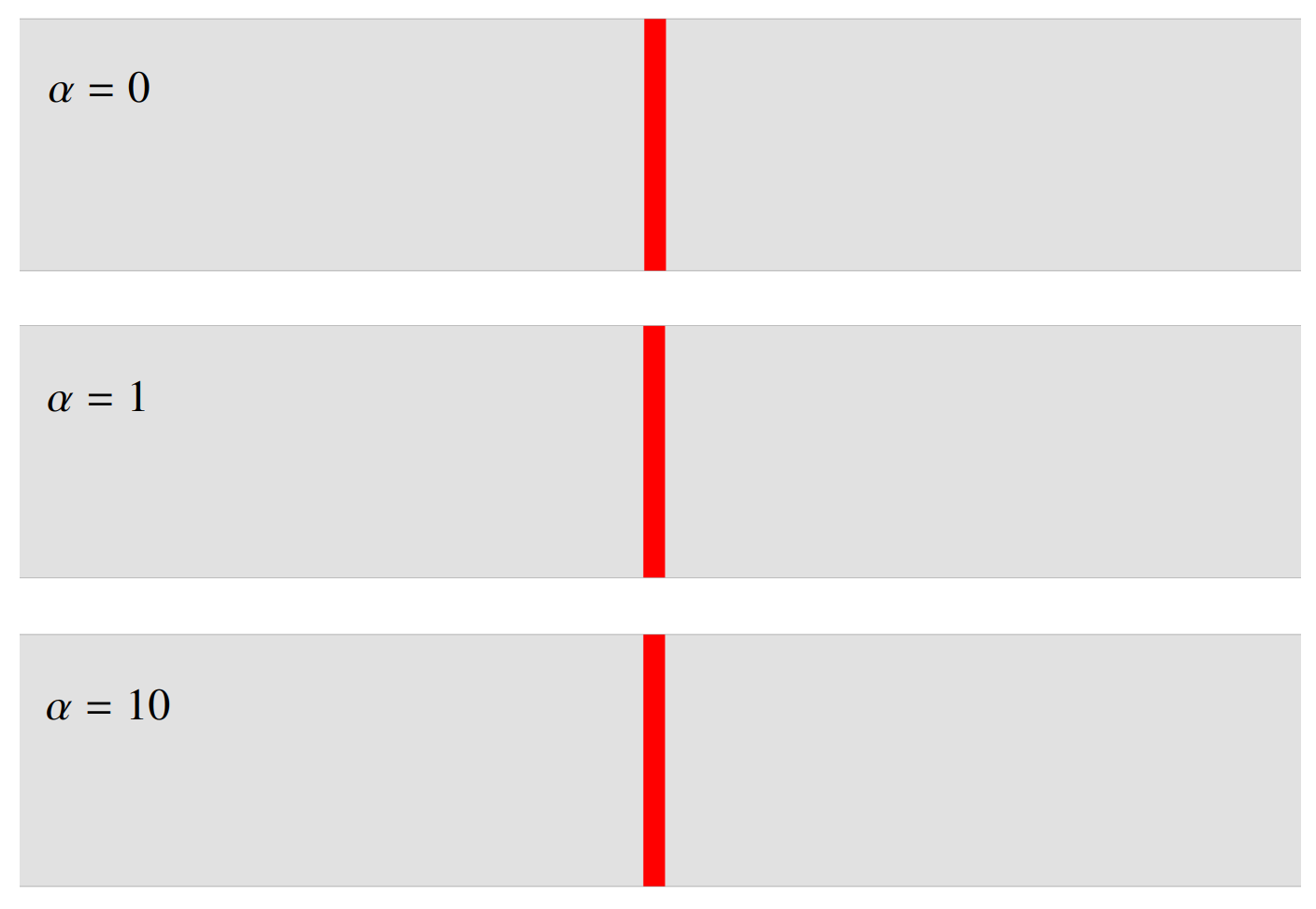}
      \caption{Crack surfaces for preferred cracking direction $\theta = 0^\circ$ and different values of $\alpha$.}
      \label{subfig:OverlayUD_00deg}
    \end{subfigure}
    \begin{subfigure}{0.49\textwidth}
      \centering
      \includegraphics[width=1.0\textwidth]{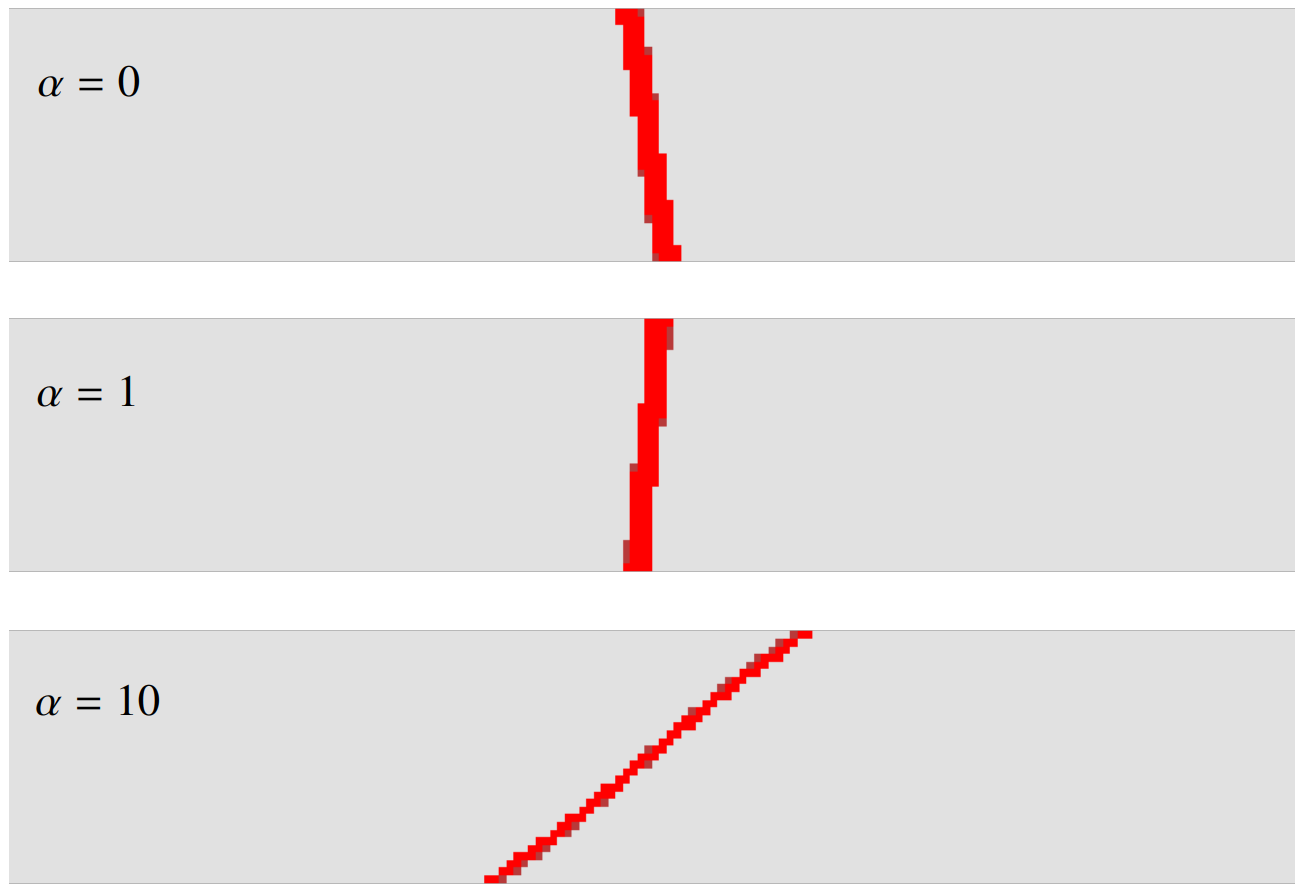}
      \caption{Crack surfaces for preferred cracking direction $\theta = 15^\circ$ and different values of $\alpha$.}
      \label{subfig:OverlayUD_15deg}
    \end{subfigure}
    \hfill
    \begin{subfigure}{0.49\textwidth}
      \centering
      \includegraphics[width=1.0\textwidth]{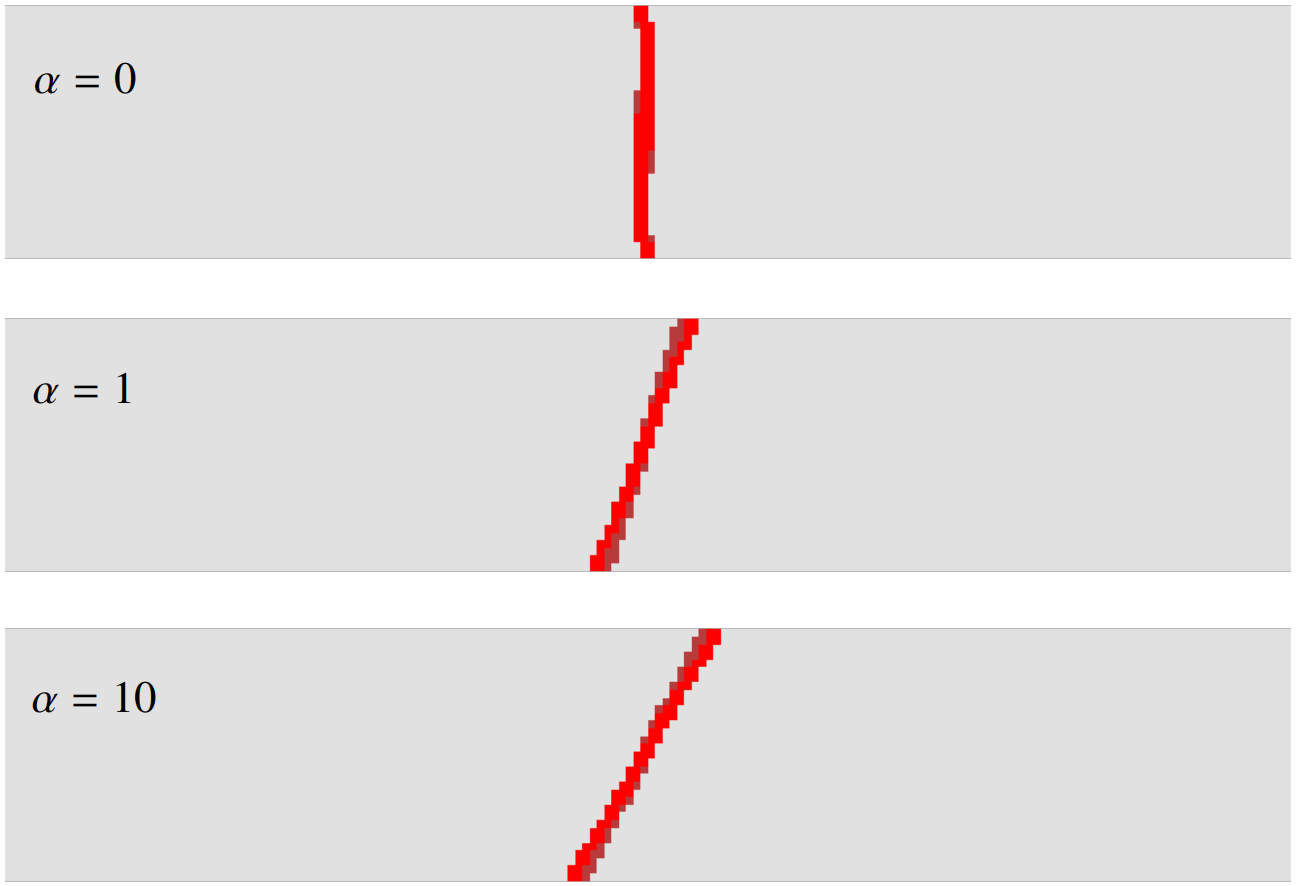}
      \caption{Crack surfaces for preferred cracking direction $\theta = 60^\circ$ and different values of $\alpha$.}
      \label{subfig:OverlayUD_60deg}
    \end{subfigure}
  \end{minipage}
    \caption{\textit{Three-dimensional uniaxial tension test.} Crack pattern for $\mathbf{a}=\bar{\Be}_1$  with $\bar{\Be}_1$ at a variable angle $\theta$ with respect to the horizontal.}
    \label{Figure:1DBar_Results_alpha}
\end{figure}

Figure~\ref{Figure:1DBar_Results_alpha} illustrates the normalized stress $\tilde{\sigma}_{11} = \sigma_{11}/\sigma_{cr}(\theta = 0^\circ,\alpha = 0)$ versus the applied displacement for equal cohesive length scales $\ell_{c_1} = \ell_{c_2} = \ell_{c_3} = 1.0$~mm. The crack orientation intensity factor $\alpha$ has no impact on the  stress-displacement response, as the response prior to nucleation does not depend on the structural tensor $\boldsymbol{A}$ and, upon nucleation, abrupt fracture occurs, see Figure~\ref{subfig:Stress_Strain_UDBar_alpha}. However, the choice of $\alpha$  has a significant effect on the direction of crack propagation. As shown in Figures~\ref{subfig:OverlayUD_00deg}, \ref{subfig:OverlayUD_15deg}, and \ref{subfig:OverlayUD_60deg}, for an increasing value of $\alpha$ the crack aligns more and more closely with the desired direction $\Ba = \bar{\Be}_1$.

\begin{figure}[h!]
   \centering
    \includegraphics[width=0.49\textwidth]{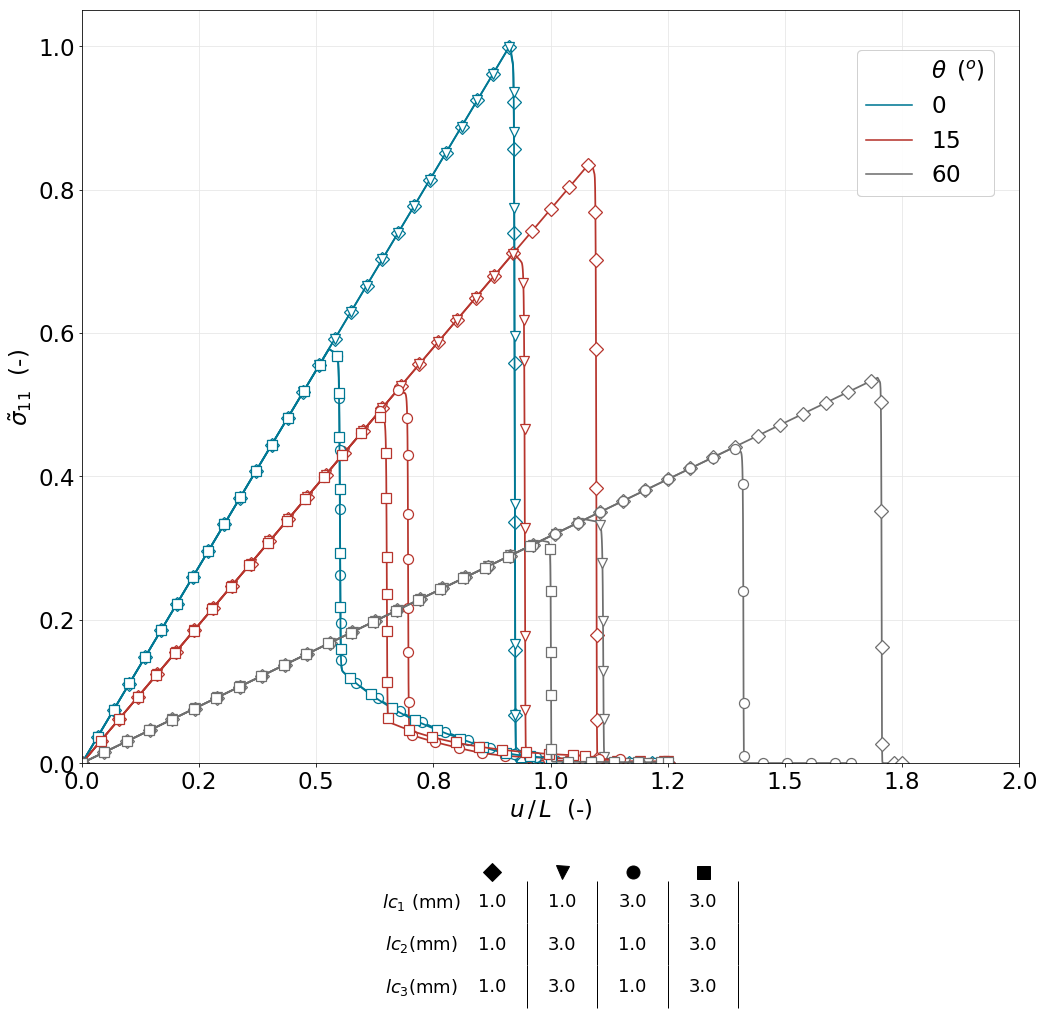}
     \caption{\textit{Three-dimensional uniaxial tension test.} Normalized stress $\tilde{\sigma}_{11} = \sigma_{11}/\sigma_{cr}(\theta = 0^\circ, \ell_{c_i} = 1$ mm$)$ versus imposed displacement for different material orientations $\theta$ and cohesive lengths $\ell_{c_i}$.}
     \label{Figure:Stress_Strain_UDBar_lci}
 \end{figure}

Next, we investigate the effect of the cohesive lengths. For three material orientations ($\theta = 0^\circ, 15^\circ, 60^\circ$), with
fixed $\hat{G}_c = 1.0$~N/mm$^2$ and  $\alpha = 10.0$, 
we examine how variations in  $\ell_{c_i}$ influence the resulting stress-displacement behavior, as illustrated in Figure~\ref{Figure:Stress_Strain_UDBar_lci}.
The effect of the cohesive lengths on the critical stress is the same already demonstrated in Section \ref{subsubsct:GoverningEqNewModel}. For instance, the ratio of the critical stresses for the cohesive length combinations $\ell_{c_1} = \ell_{c_2} = \ell_{c_3} = 1.0$~mm and $\ell_{c_1} = 3.0$~mm, $\ell_{c_2} = \ell_{c_3} = 1.0$~mm varies for different angles (it is largest for $\theta = 0^\circ$ and smallest for $\theta = 60^\circ$). 
This demonstrates the flexibility of the proposed model to independently control the critical stress across different orientations,  making it suitable to describe realistic fracture experimental data in anisotropic materials. Moreover, Figure~\ref{Figure:Stress_Strain_UDBar_lci} shows in some cases the presence of a cohesive response after crack nucleation.

\subsection{Plane-strain compact tension test}
\label{sct:CompactTensionTest}
\begin{figure}[h!]
  \centering
  \includegraphics[width=0.65\textwidth]{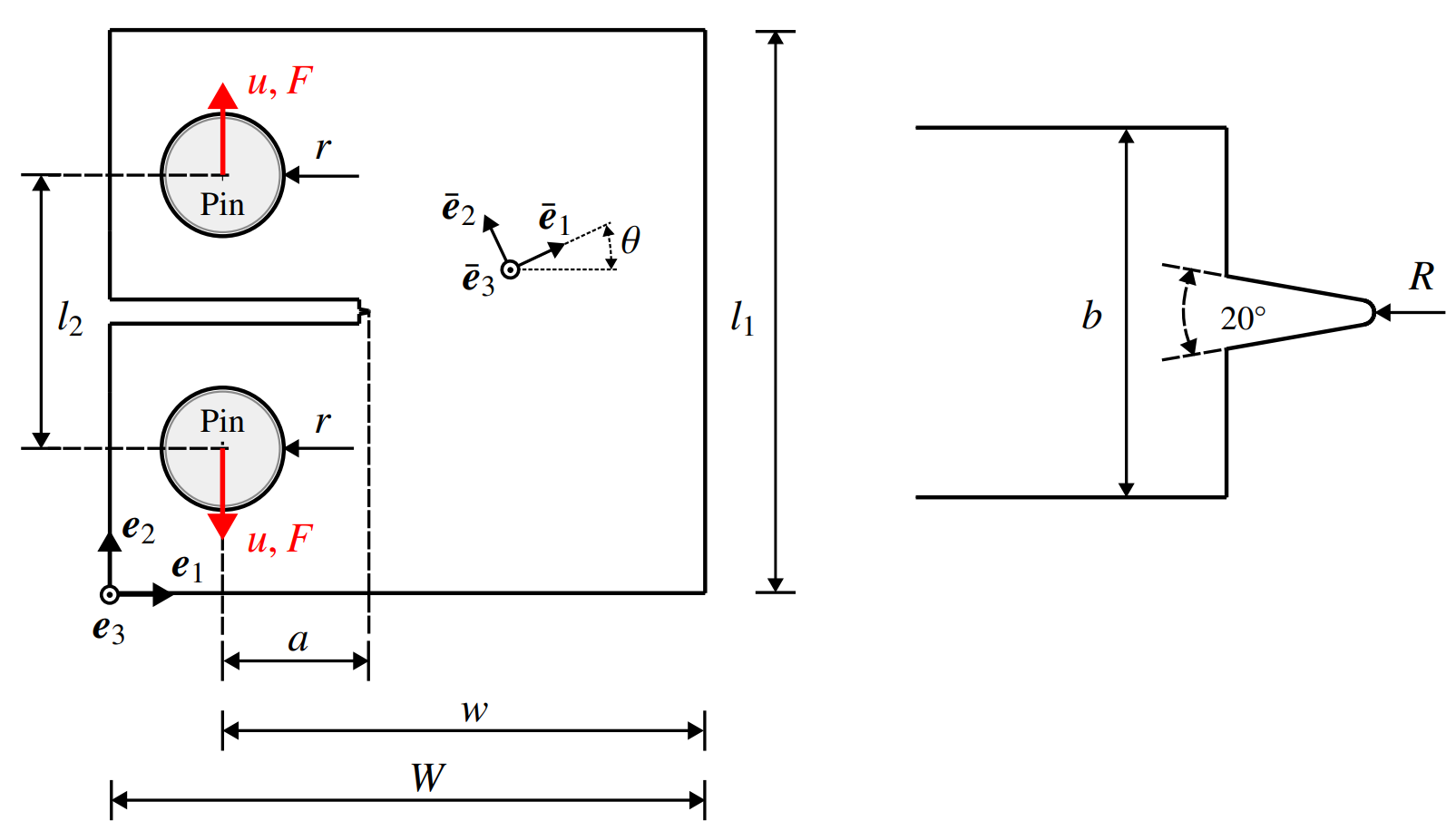}
  \caption{\textit{Plane-strain compact tension test.} The sample has dimensions $l_1 = 7$~mm, $l_2 = 3.4$~mm, $w_1 = 6$~mm, $w_2 = 7.4$~mm, $a = 1.4$~mm, $r = 0.75$~mm, $b = 0.3$~mm and $R =0.01$~mm, and is subjected to an imposed displacement $u = 1$~mm applied in increments $\Delta u = 2.5 \cdot 10^{-4}$~mm via rigid pins inserted into the  holes in contact with the sample. The domain is meshed using 8-node brick elements with a global seed size of $l_h = 0.008$~mm.}
  \label{Figure:Sketch_ThreeDimensionalCTBVP}
\end{figure}

The compact tension sample in Figure~\ref{Figure:Sketch_ThreeDimensionalCTBVP} is pulled by two rigid pins  inserted into the holes, for which we model contact with the corresponding surfaces of the sample.
The plane-strain conditions are realized by a discretization with one element over the thickness and constrained out-of-plane displacements. The anisotropic material behavior is defined in the local coordinate system $\bar{\Be}_i$ at an angle $\theta$ with respect to the global coordinate system $\Be_i$.
The aim of this example is to demonstrate the capability of the model to induce crack propagation along a desired material direction in presence of an initial stress singularity.

\begin{figure}[h!]
    \centering
    \begin{subfigure}[b]{0.49\textwidth}
        \centering
        \includegraphics[width=\textwidth]{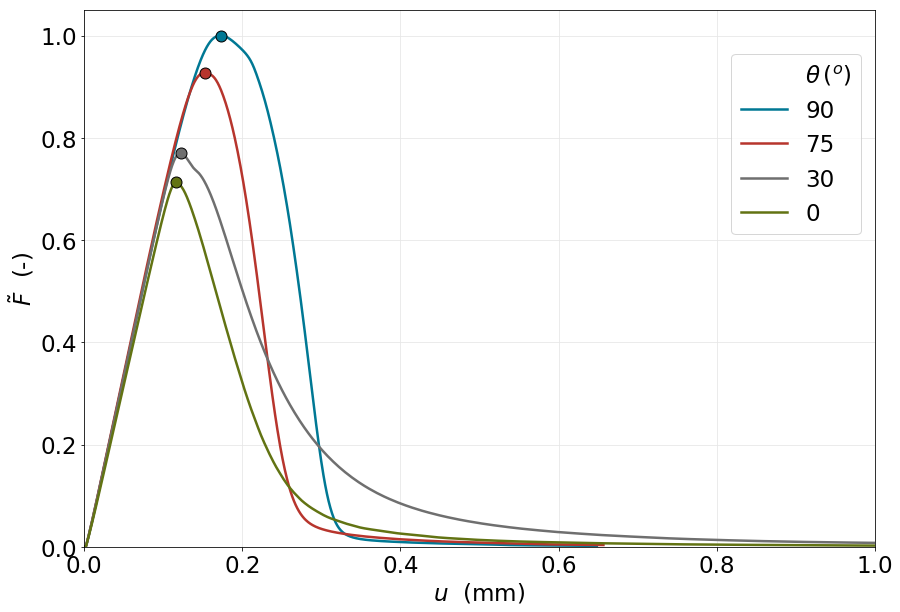}
        \caption{Normalized force-displacement response with $\tilde{F} = F(\theta)/F_{max}(\theta=0^\circ)$.}
        \label{subfig:Results_CT_ForceDisplacement}
    \end{subfigure}
    \hfill
    \begin{subfigure}[b]{0.49\textwidth}
        \centering
        \includegraphics[width=\textwidth]{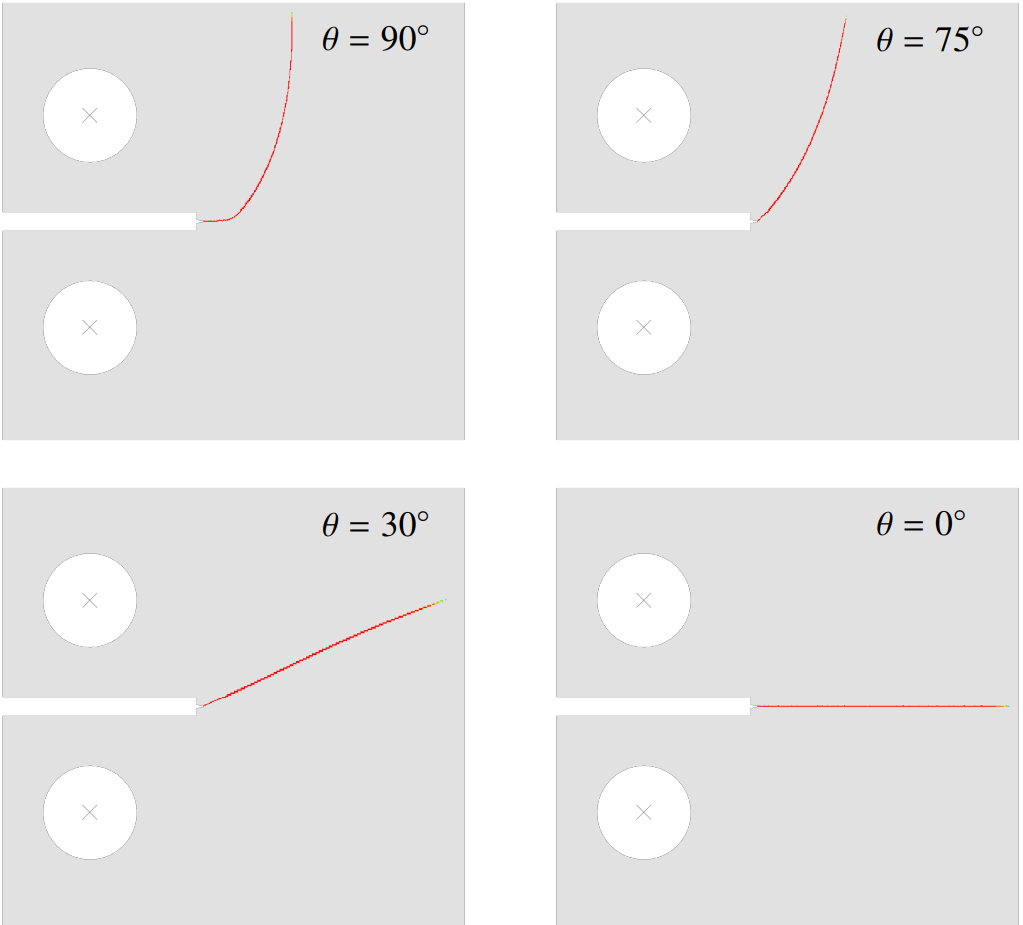}
        \caption{Resulting crack topologies for a variation of the local material orientation $\theta$.}
        \label{subfig:Results_CT_OverlayPlots}
    \end{subfigure}
    \caption{\textit{Plane-strain compact tension test.} Structural response and crack surfaces for varying material orientation.}
    \label{Figure:Results_CT2D}
\end{figure}

Figure \ref{Figure:Results_CT2D} illustrates the main results for the material orientations $\theta \!=\! 90^\circ$, $\theta \!=\! 75^\circ$, $\theta \!=\! 30^\circ$, and $\theta = 0^\circ$. Figure~\ref{subfig:Results_CT_ForceDisplacement} depicts the reaction force versus the applied displacement; as expected, the  peak load varies with the material orientation $\theta$. 
Figure~\ref{subfig:Results_CT_OverlayPlots} highlights the ability of the model to induce cracking in the desired material direction $\Ba = \bar{\Be}_1$. For $\theta = 0^\circ$, the crack follows the horizontal direction $\Ba = \bar{\Be}_1 = \Be_1$ from the very beginning of the fracture process, whereas for the extreme case $\theta = 90^\circ$ it tilts from the initial $0^\circ$ direction towards the desired $\Ba = \bar{\Be}_1 = \Be_2$ direction.
%
%

\subsection{Plane-strain single-edge notched tension test with layered structure}
\label{subsct:SPBTest}
\begin{figure}[h!]
  \centering
  \includegraphics[width=0.65\textwidth]{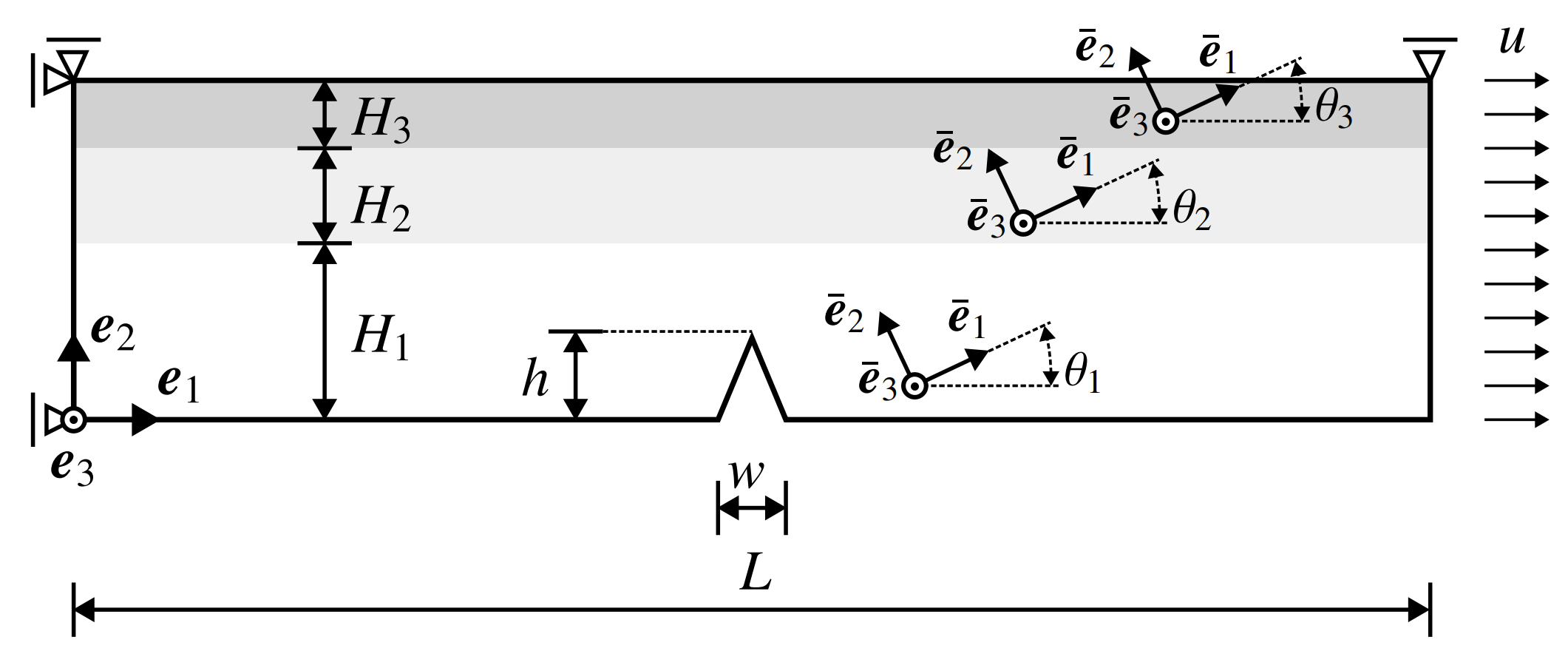}
  \caption{\textit{Plain strain single-edge notched tension test with layered structure.} The sample has dimensions $L = 100$~mm, $H_1 = 13$~mm, $H_2 = 7$~mm,  $H_3 = 5$~mm, $w = 5$~mm and $h = 6$~mm, with layered local orientations $\underline{\theta} = [\theta_1, \theta_2, \theta_3]$, and is subjected to an imposed displacement $u = 1$~mm applied in increments $\Delta u = 5 \cdot 10^{-4}$~mm. The domain is meshed using 8-node brick elements with global seed size of $l_h = 0.04$~mm.}
  \label{Figure:Sketch_SENTBPV_2D}
\end{figure}

To assess the ability of the model to reproduce changes in the direction of crack propagation when the material weak direction changes abruptly, we investigate a plane-strain single-edge notched tension test on a sample with layered geometry, see Figure \ref{Figure:Sketch_SENTBPV_2D}. Plane-strain conditions are realized by a mesh with one element over the thickness where the out-of-plane displacements at both surfaces are constrained. Each layer of the sample has a local material orientation $\bar{\Be}_i$ at an angle $\theta_i$ with respect to the global coordinate system.
We investigate three different configurations of the layered structure, namely $\underline{\theta} = [0^\circ, 30^\circ, 0^\circ]$, $\underline{\theta} = [45^\circ, 0^\circ, 30^\circ]$, and $\underline{\theta} = [60^\circ, 45^\circ, 0^\circ]$. 
\begin{figure}[h!]
  \centering
  \includegraphics[width=0.65\textwidth]{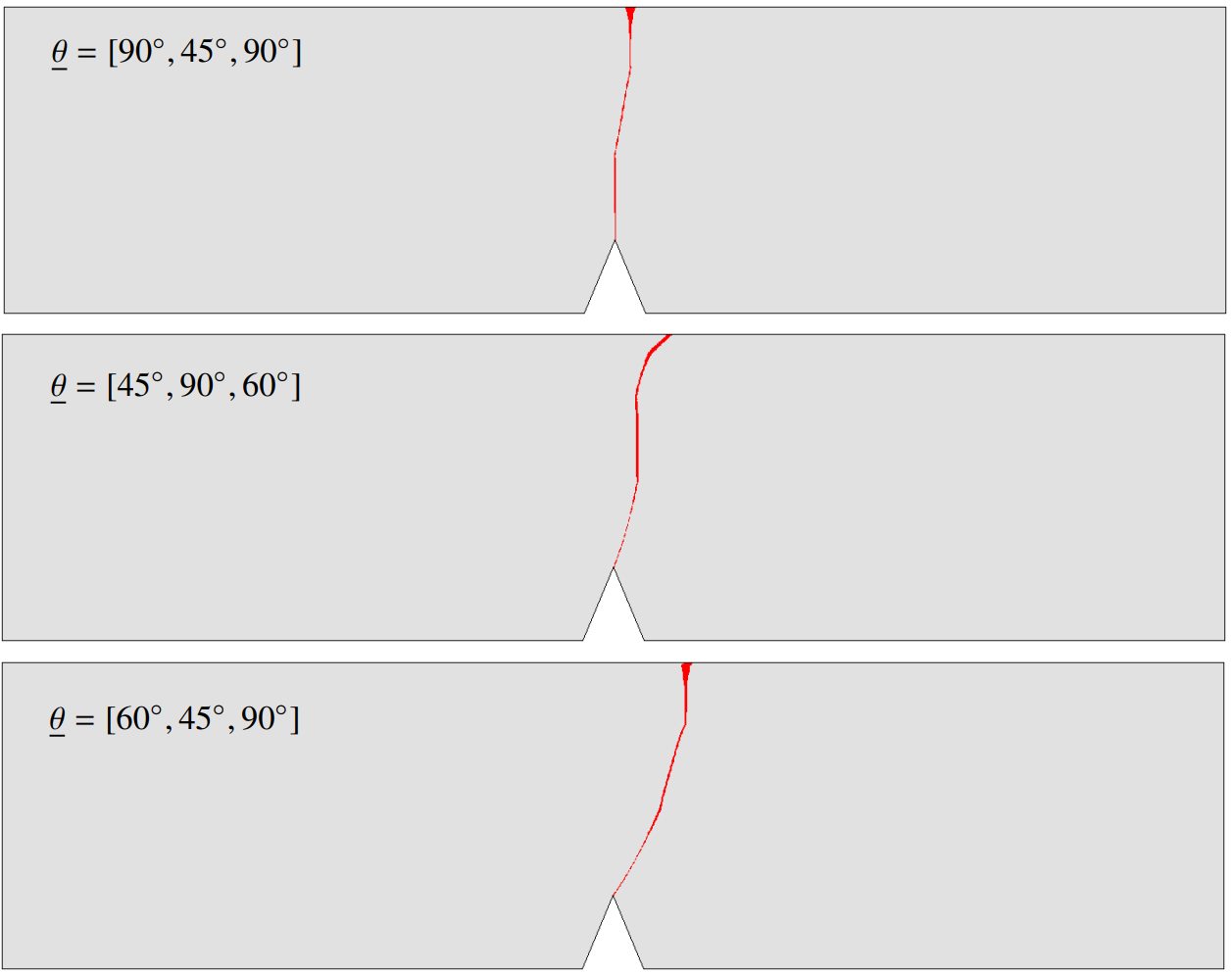}
  \caption{\textit{Plain-strain single-edge notched tension test with layered structure.} Crack surfaces for different layer setups.}
  \label{Figure:ResultsSENT_2D}
\end{figure}
The crack patterns in Figure \ref{Figure:ResultsSENT_2D} demonstrate the ability of the model to deflect the crack propagation angle along the weak material direction when this direction changes from layer to layer.

\subsection{Three-dimensional single-edge notched tension test with laminate structure}
\label{subsct:BendingTest3D}
\begin{figure}[h!]
  \centering
  \includegraphics[width=0.8\textwidth]{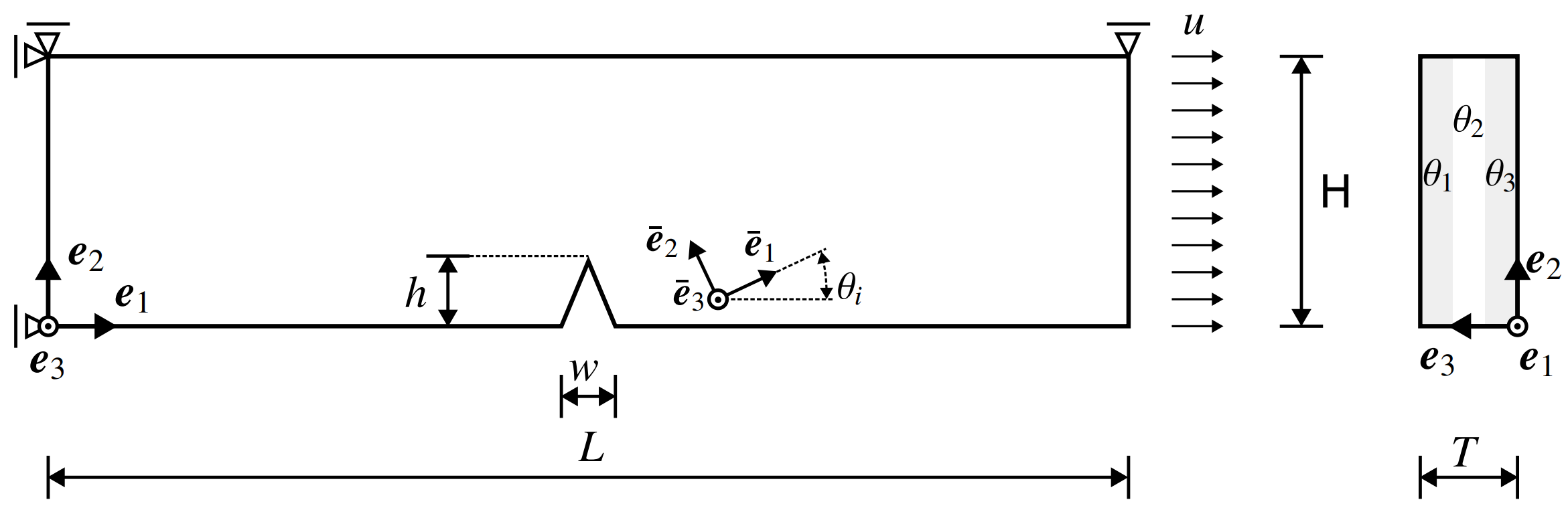}
  \caption{\textit{Three-dimensional single-edge notched tension test with laminate structure.} The sample has dimensions $L = 100$~mm, $H = 25$~mm, and $T = 9$~mm, $w = 5$~mm and $h = 6$~mm, and layered local orientations $\underline{\theta} = [\theta_1, \theta_2, \theta_3]$, and is subjected to an imposed displacement $u =2.5$~mm applied in increments $\Delta u = 1.25 \cdot 10^{-3}$~mm. The domain is meshed using 8-node brick elements with a global seed size of $l_h = 0.25$~mm.}
  \label{Figure:Sketch_SENTBPV_3D}
\end{figure}

To investigate a more complex fracture situation with 3D effects, inspired by Nguyen and Waas~\cite{Nguyen2022}, we investigate a 3D single-edge notched tension test with an out-of-plane layered structure, see Figure~\ref{Figure:Sketch_SENTBPV_3D}. In the thickness direction, the sample has a laminate structure with three  layers of equal thickness but different local coordinate systems $\bar{\Be}_i$. 
%
We investigate four laminate configurations, i.e., $\underline{\theta} = [30^\circ, 120^\circ, 30^\circ]$, $\underline{\theta} = [120^\circ, 30^\circ, 120^\circ]$, $\underline{\theta} = [45^\circ, 90^\circ, 45^\circ]$, and $\underline{\theta} = [90^\circ, 45^\circ, 90^\circ]$.
%
%
%
\begin{figure}[h!]
   \centering
   \begin{subfigure}[b]{0.245\textwidth}
       \centering
        \includegraphics[scale=0.45]{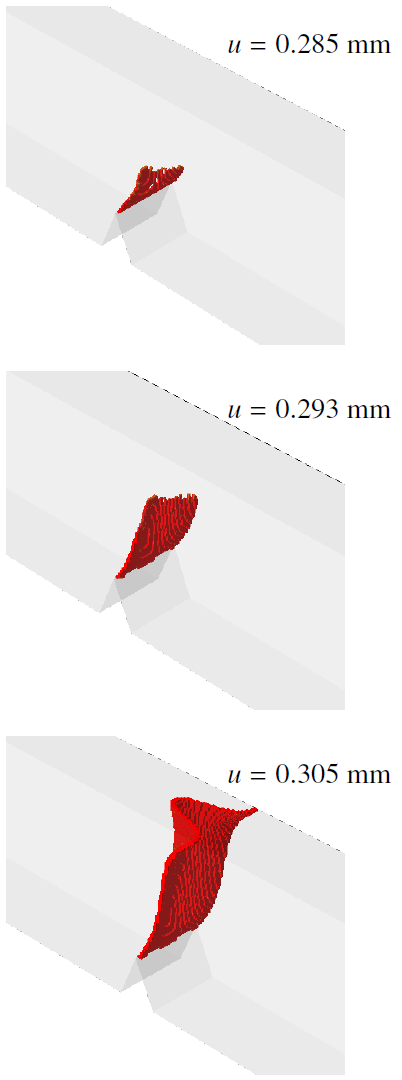}
        \caption{Setup $\underline{\theta} = [30^\circ, 120^\circ, 30^\circ]$}
        \label{subfig:SENT_3D_Sequence_30_120_30}
    \end{subfigure}
    \begin{subfigure}[b]{0.245\textwidth}
       \centering
        \includegraphics[scale=0.45]{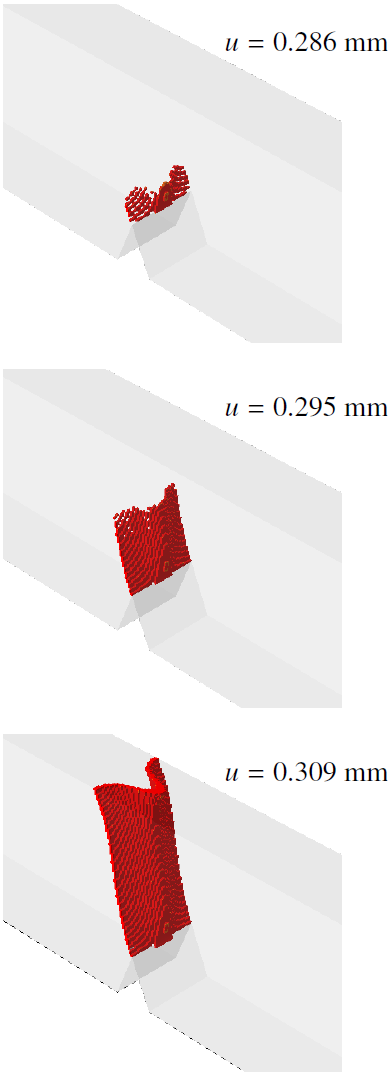}
        \caption{Setup $\underline{\theta} = [120^\circ, 30^\circ, 120^\circ]$}
        \label{subfig:SENT_3D_Sequence_120_30_120}
    \end{subfigure}
    \begin{subfigure}[b]{0.245\textwidth}
       \centering
        \includegraphics[scale=0.45]{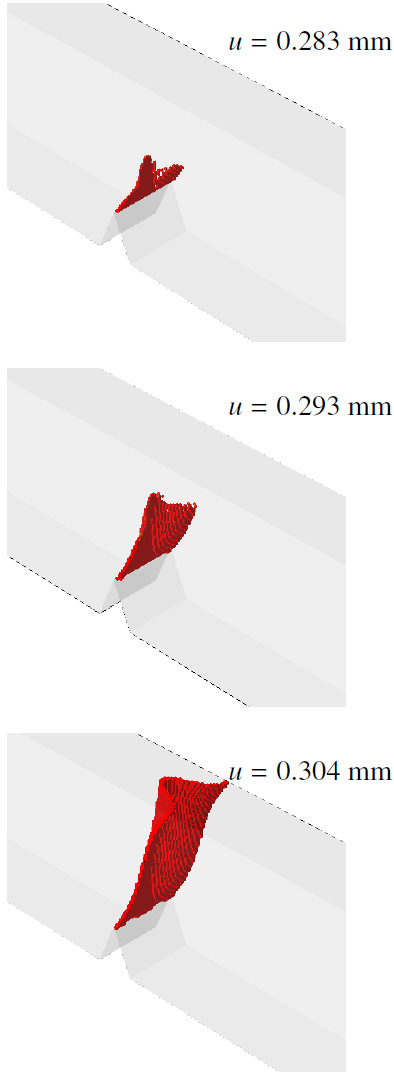}
        \caption{Setup $\underline{\theta} = [45^\circ, 90^\circ, 45^\circ]$}
        \label{subfig:SENT_3D_Sequence_45_90_45}
    \end{subfigure}
    \begin{subfigure}[b]{0.245\textwidth}
       \centering
        \includegraphics[scale=0.45]{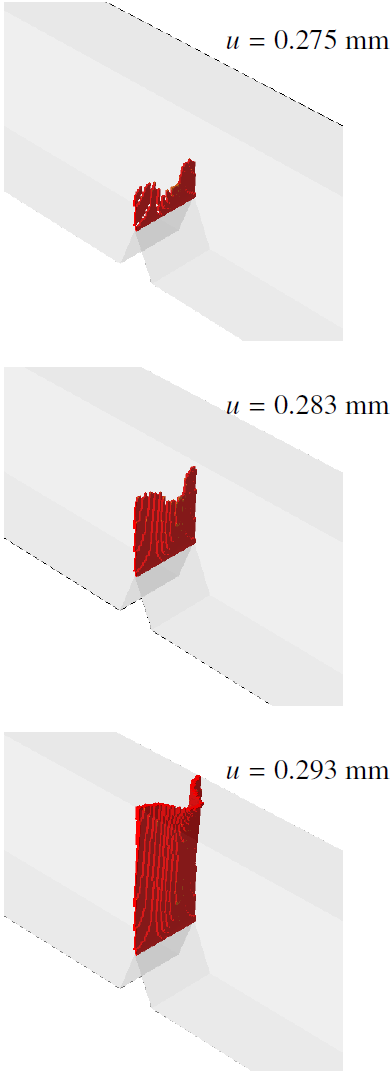}
        \caption{Setup $\underline{\theta} = [90^\circ, 45^\circ, 90^\circ]$}
        \label{subfig:SENT_3D_Sequence_90_45_90}
    \end{subfigure}
    \caption{\textit{Three-dimensional single-edge notched tension test with laminate structure.} Crack surfaces at different applied displacements for different composite layouts. The damage threshold for visualization is chosen as $d=0.95$.}
    \label{Figure:SENT3D_Sequence}
\end{figure}
%

Figure~\ref{Figure:SENT3D_Sequence} shows snapshots of crack propagation in samples with the different laminate layouts, yielding complex crack surfaces in the 3D space. In Figures~\ref{subfig:SENT_3D_Sequence_30_120_30} and \ref{subfig:SENT_3D_Sequence_120_30_120}, the crack surface is tilted along the orientations $\theta = 30^\circ$ or $\theta = -30^\circ$ of the outer layers, and the orientation of the middle layer leads to a U-shaped crack. Similarly, in Figures~\ref{subfig:SENT_3D_Sequence_45_90_45} and \ref{subfig:SENT_3D_Sequence_90_45_90} the orientation of the crack  is dominated by that of the outer layers. However, the crack grows in the middle, accounting for the inner layer alignment and causing again a U-shape. 

\section{Conclusions}
\label{sct:Conclusion}
We presented a novel variational phase-field model for fracture in anisotropic materials. The proposed \textit{multi-cohesive} model enhances the flexibility of existing anisotropic formulations in capturing nucleation of a new crack by introducing an asymptotically cohesive degradation function with distinct cohesive lengths along the principal material directions, while preserving a single damage variable. Based on analytical investigations verified numerically, we concluded that with the proposed model the strength surface under multiaxial stress states (or alternatively the directional critical stress under uniaxial tension) can be more flexibly calibrated than with the standard anisotropic model. Also, in contrast to the model with multiple damage variables, the proposed model leads to a smooth strength surface, which may be more representative of the multiaxial strength response of at least some anisotropic materials. Further, for the proposed model we studied the role of elastic and cohesive fracture anisotropy on the second-order stability of the homogeneous solution at the onset of damage. We obtained that, under purely volumetric stress states, anisotropy tends to promote instability of the homogeneous solution; conversely, under uniaxial stress states, since these are already quite far from being purely volumetric, anisotropy may stabilize the homogeneous solution. Consistently with the results of previous investigations on cohesive phase-field fracture models for isotropic solids, we concluded that increasing the ratios between cohesive lengths and regularization length promotes the stability of the homogeneous solution. Finally, numerical simulations in two and three dimensions demonstrated the ability of the model to reproduce complex fracture phenomena, including directional crack nucleation, propagation along preferred material orientations, and deflection across layered or laminated configurations. Overall, the proposed approach combines the computational efficiency of single-variable formulations with the directional flexibility achievable through multi-variable models, thereby offering an attractive framework for modeling anisotropic fracture.

\clearpage

\appendix
\section{Second-order stability of the homogeneous solution: 1D analysis}
\label{sct:Appendix}
In this appendix, we recall the 1D results of~\cite{Pham_2010,Pham2011a} on the stability of the homogeneous solution and apply them to the gradient damage model of~\cite{Lorentz_2010,Lorentz2012} with the AT$_1$ local dissipation function \eqref{eq:wd_SM}. 
We consider a bar \(\Omega = (0, L)\) with Dirichlet boundary conditions \(u(0)=0\), \(u(L)=u_D\). 
The material is isotropic, characterized by the initial elastic modulus \(E_0\) and the fracture toughness \(G_c\). 
The total energy reads
\begin{equation}
    \calE(u,d) = \int_0^L \left\{ \tfrac{1}{2} g(d)\,E_0\,u'^2 + \tfrac{3G_c}{8\ell} \left( d + \ell^2 d'^2 \right) \right\} dx,
    \label{eq:OneDimensionalFunctional}
\end{equation}
with degradation function
\begin{equation}
    g(d) = \frac{(1-d)^2}{(1-d)^2 + 2r\,d\,(1+pd)}, \quad r=\tfrac{\ell_c}{\ell},
    \label{eq:lorentz1D}
\end{equation}
where  $\ell_c$ is the cohesive length in the 1D setting. 

The homogeneous solution corresponds to the first-order stable state with vanishing damage gradient, $d'=0$. It is therefore
\begin{equation}
u(x) = \tfrac{u_D}{L}\,x \quad \text{and} \quad
d = \begin{cases}
0, & \text{if } u_D \leq {u_D}_e=\tfrac{1}{2}\sqrt{\tfrac{3}{2r}\,\tfrac{G_c}{E_0\ell}}\,L,\\[1mm]
f^{-1}\!\left(-\tfrac{3}{4}\,\tfrac{G_c}{E_0\ell}\,\tfrac{L^2}{u_D^2}\right), & \text{otherwise},
\end{cases}
\label{eq:hom_sol}
\end{equation}
where $f(d)=g'(d)$.

The time-continuous second-order stability analysis in~\cite{Pham_2010,Pham2011a} yields a general criterion for gradient damage models. 
Applied to \eqref{eq:OneDimensionalFunctional}--\eqref{eq:lorentz1D}, it shows that the homogeneous solution \eqref{eq:hom_sol} is stable provided $L \leq L_s(u_D)$, with
\begin{equation}
\begin{aligned}
        L_s(u_D) = &
    2\,\sqrt{\frac{G_c\,r}{c_w\,\ell\,E_0}}\,\frac{\left(1+(1+2\,p)\,d\right)^2}{\sqrt{\left(2+p+(1+2\,p)\,d\right)^3}}\,\frac{\pi\,L}{u_D}\\
    =&\sqrt{\frac{3}{2}}\,\sqrt{\frac{G_c\,r\,\ell}{E_0}}\,\frac{\left(1+(1+2\,p)\,f^{-1}\!\left(-\tfrac{3}{4}\,\tfrac{G_c}{E_0\ell}\,\tfrac{L^2}{u_D^2}\right)\right)^2}{\sqrt{\left(2+p+(1+2\,p)\,f^{-1}\!\left(-\tfrac{3}{4}\,\tfrac{G_c}{E_0\ell}\,\tfrac{L^2}{u_D^2}\right)\right)^3}}\,\frac{\pi\,L}{u_D}.
\end{aligned}
\label{eq:CritLs}
\end{equation}

To assess the stability of the homogeneous solution at the onset of damage ($d \to 0^+$), we consider
\begin{equation}
    L_s({u_D}_e) = \frac{2 \pi}{(2+p)^{3/2}} \, r \, \ell = \frac{2 \pi}{(2+p)^{3/2}} \, \ell_c.
\end{equation}
This expression is similar to the state-stability result in equation (41) of \cite{zolesi2024stability}. In particular, from this we obtain that the homogeneous solution is stable if
\begin{equation}
    \frac{2\pi}{(2+p)^{3/2}}\,r\,\frac{\ell}{L}\geq 1,
\end{equation}
Thus, we see that for a large domain, i.e., \(L/\ell \rightarrow \infty\), the 1D homogeneous solution is unstable. The same conclusion does not generally hold for the higher-dimensional case (\(n_{\text{dim}} > 1\)) discussed in \ref{AppB}, where the stability of the homogeneous solution in large structures depends on the stress state and on the ratio between the cohesive and regularization lengths.

\section{Second-order stability of the homogeneous solution: vectorial analysis}
\label{AppB}
As follows, we apply the analytical results of \cite{pham2013stability} to derive a stability criterion for the homogeneous solution of the MCM with the AT$_1$ linear dissipation function \eqref{eq:wd_SM} in the vectorial case (\(n_{\rm dim} > 1\)), under the simplifying assumption of a sufficiently large domain \(\Omega\), i.e., such that the characteristic size \(L\) of the domain satisfies \(L/\ell \rightarrow \infty\).
The homogeneous solution is given by $(\bar{\boldsymbol{u}} = \bar{\boldsymbol{\varepsilon}}\,\boldsymbol{x}, \bar{d})$, which satisfies the first-order stability condition \eqref{eq:NecessaryCondition}, with $\bar{\boldsymbol{\varepsilon}}$ and $\bar{d}$ constant in $\Omega$. Consequently, the associated stress field $\bar{\boldsymbol{\sigma}} = \mathbb{C}(\bar{d})\,\bar{\boldsymbol{\varepsilon}}$ is also homogeneous.  

For a unit vector $\boldsymbol{k}\in \mathbb{R}^{n_{dim}}$, we introduce the functional space  
\begin{equation}
    \mathcal{M}(\boldsymbol{k})=\bigg\{\,\boldsymbol{k}\otimes\boldsymbol{v}+\boldsymbol{v}\otimes\boldsymbol{k}\;:\;\boldsymbol{v}\in\mathbb{R}^{n_{dim}}\bigg\},
    \label{eq:jump_compatible_space}
\end{equation}
which we call the \textit{jump compatible} space, echoing the terminology introduced in the context of phase-field modeling of cohesive fracture in \cite{bourdin2025variational}.
According to the second-order analysis in \cite{pham2013stability}, the homogeneous solution $(\bar{\boldsymbol{u}}, \bar{d})$ is stable only if  
\begin{equation}
\label{eq:Rayleigh_ratio}
    \mathsf{R}=\frac{ 
    \boldsymbol{q}(\bar{d}):\mathbb{C}(\bar{d}):\boldsymbol{q}(\bar{d}) - \mathsf{V}}
    {\tfrac{1}{2}\,\bar{\boldsymbol{\sigma}}:\mathbb{S}''(\bar{d}):\bar{\boldsymbol{\sigma}}}\geq 1,\quad\text{with}\quad   \boldsymbol{q}(d)=\mathbb{S}'(\bar{d}):\bar{\boldsymbol{\sigma}}
\end{equation}
and  
\begin{equation}
\mathsf{V}= \max_{\{\boldsymbol{k}: \vert\boldsymbol{k}\vert=1\}} 
    \mathbb{C}(\bar{d})\,\boldsymbol{\eta}(\boldsymbol{k})\cdot \boldsymbol{\eta}(\boldsymbol{k}),
    \label{eq:max_V}
\end{equation}
where $\boldsymbol{\eta}(\boldsymbol{k})\in\mathcal{M}(\boldsymbol{k})$ is the minimizer  
\begin{equation}
    \boldsymbol{\eta}(\boldsymbol{k})
    =\argmin_{\boldsymbol{\eta}\in\mathcal{M}(\boldsymbol{k})}
    \mathbb{C}(\bar{d})\big(\boldsymbol{q}(\bar{d})-\boldsymbol{\eta}\big)\cdot 
    \big(\boldsymbol{q}(\bar{d})-\boldsymbol{\eta}\big).
    \label{eq:min_eta}
\end{equation}
The variational problem \eqref{eq:min_eta} represents the minimization of a distance weighted by the elasticity tensor \(\mathbb{C}(\bar{d})\). Accordingly, \(\boldsymbol{\eta}(\boldsymbol{k})\) can be interpreted as the projection of \(\boldsymbol{q}(\bar{d})\) onto the jump compatibility space \(\mathcal{M}\) \cite{kohn1991relaxation}.

According to Proposition 3 in \cite{pham2013stability}, the Rayleigh ratio \(\mathsf{R} = 0\), i.e., $\boldsymbol{q}(\bar{d})\in\mathcal{M}(\boldsymbol{k})$ and \(\boldsymbol{\eta}(\boldsymbol{k}) = \boldsymbol{q}(\bar{d})\), if and only if the symmetric tensor \(\boldsymbol{q}(\bar{d})\) is
\begin{enumerate}
    \item a rank-one tensor, or
    \item a rank-two tensor and its non-zero eigenvalues have opposite signs.
\end{enumerate}
Thus, in these cases, the homogeneous solution is unstable. 
If these conditions on \(\boldsymbol{q}(d)\) are not satisfied, the Rayleigh ratio is greater than zero, 
and \(\mathsf{R}\) must be computed to determine whether the homogeneous solution is stable or unstable. 
For the numerical calculations of \(\mathsf{R}\), as in \cite{zolesi2024stability}, we consider a purely 2D problem. This choice simplifies the computations compared to the 3D case, while still capturing the same qualitative features of the stability analysis, which cannot be observed in a 1D study. Since we analyze a 2D setting, the vectors \(\boldsymbol{v}\) and \(\boldsymbol{k}\) are defined as \(\boldsymbol{v}=[v_1,v_2]^T\) and \(\boldsymbol{k}=[k_1,k_2]^T\), respectively.
We focus on the limit case \(d \to 0^{+}\), which represents the most relevant scenario for preventing a stable homogeneous solution and for allowing the localization of cracks from the onset of damage. 
Algorithmically, we first define the matrix \(\underline{\boldsymbol{\eta}}(\boldsymbol{k})\) in Voigt notation as

\begin{equation}
    \underline{\boldsymbol{\eta}}(\boldsymbol{v},\boldsymbol{k}) = 2 \begin{bmatrix}
        v_1 k_1\\
        v_2 k_2\\
        v_1 k_2 + v_2 k_1
    \end{bmatrix}.
\end{equation}
Then, the minimization problem \eqref{eq:min_eta} can be reformulated as the linear problem of finding \(\boldsymbol{v}(\boldsymbol{k}) = [v_1(\boldsymbol{k}), v_2(\boldsymbol{k})]^T\) such that

\begin{equation} 
\nabla \mathcal{H}_{\boldsymbol{k}}(v_1(\boldsymbol{k}),v_2(\boldsymbol{k})) = \boldsymbol{0}, \quad \text{with}\quad \mathcal{H}_{\boldsymbol{k}}(v_1,v_2) = \left(\underline{\boldsymbol{q}}(0)-\underline{\boldsymbol{\eta}}(\boldsymbol{k})\right)^T\,\underline{\mathbb{C}}^{2D}_0\,\left(\underline{\boldsymbol{q}}(0)-\underline{\boldsymbol{\eta}}(\boldsymbol{k})\right), 
\end{equation}
where \(\underline{\boldsymbol{q}}(d)=[q_{11}(d), q_{22}(d), 2q_{12}]^T\) constitutes the tensor \(\boldsymbol{q}(d)\) in Voigt notation. We then introduce
\begin{equation}
    b(\omega) = \mathcal{B}\left([\cos(\omega), \sin(\omega)]^T\right), \quad \text{with}\quad \omega \in [0,2\pi],
\end{equation}
where
\begin{equation}
    \mathcal{B}(\boldsymbol{k}) = \underline{\boldsymbol{\eta}}(\boldsymbol{v}(\boldsymbol{k}),\boldsymbol{k})^T 
    \,\underline{\mathbb{C}}^{2D}_0\, 
    \underline{\boldsymbol{\eta}}(\boldsymbol{v}(\boldsymbol{k}),\boldsymbol{k}).
\end{equation}
Thus, according to \eqref{eq:max_V}, the value \(\mathsf{V}\) is obtained by solving the single-variable maximization problem
\begin{equation}
    \mathsf{V} = \max_{\omega \in [0,2\pi]} b(\omega),
\end{equation}
from which the Rayleigh ratio \(\mathsf{R}\) can be computed straightforwardly using \eqref{eq:Rayleigh_ratio}.

In the following, we consider an elastic orthotropic material with 
\begin{equation}
    \underline{\mathbb{S}}^{2D}_0=\begin{bmatrix}
        \frac{1}{E_1} &-\frac{\nu_{21}}{E_2} & 0\\
        -\frac{\nu_{21}}{E_2} & \frac{1}{E_2} & 0\\  
        0 & 0 & \frac{1}{G_{12}}
    \end{bmatrix}
\end{equation}
and the parameters listed in Table~\ref{tab:parameters}. 
Note that, with the definition of \(G_{12}\) given in the table, the elastic behavior of the material is isotropic when \(k_E = 1\), whereas its damage behavior is isotropic when \(k_r = 1\).

The value of \(\mathsf{R}\) depends on \(k_E\), \(k_r\), \(r\), and the specific homogeneous stress state considered.  
We study two scenarios. First, knowing that for brittle isotropic materials the stress states dominated by the volumetric component tend to stabilize the homogeneous solution even for large domains \cite{zolesi2024stability}, we consider a purely \textit{volumetric stress state} to investigate the effect of anisotropy on stability. Second, we also consider the \textit{uniaxial stress state} to investigate how the load direction interacts with the intrinsic material anisotropy from the standpoint of stability.
\begin{table}[h!]
\centering
\renewcommand{\arraystretch}{1.2}
\begin{tabular}{cccccccccc}
\hline
\(E_1/E_2\) & \(E_2\) & \(\nu_{21}\) & \(G_{12}\) & \(r_1/r_2\) & \(r_2\) & \(p_1, p_2\) & \(G_c\) & \(\ell\) \\
\noalign{\vskip 2pt} 
\hline
\noalign{\vskip 2pt} 
\(k_E\) & 1 &  0.25 & \(\frac{E_2}{2(1+\nu_{21})}\) & \(k_r\) & \(r\) & 2 & 1 & 0.1 \\
\noalign{\vskip 2pt}
\hline
\end{tabular}
\caption{Material parameters.}
\label{tab:parameters}
\end{table}

\subsection{Volumetric stress state}
\label{AppB_vol}
We analyze a purely volumetric stress state in 2D, i.e., the $2 \times 2$ stress tensor is given by
\[
\bar{\boldsymbol{\sigma}} = p^{ cr}\,\boldsymbol{I},
\]
where the critical pressure $p^{cr}$ at the onset of damage is
\begin{equation}
    p^{cr} = \sqrt{\frac{2\,G_c}{c_w \ell\, \boldsymbol{I} : {\mathbb{S}^{2D}}'(0) : \boldsymbol{I}}} 
    = \sqrt{\frac{15}{3+\left(\frac{4}{k_E}-1\right)\,k_r}} \, \sqrt{\frac{1}{r}}.
    \label{eq:p_cr}
\end{equation}

Using the definition in \eqref{eq:Rayleigh_ratio}, we compute $\boldsymbol{q}(0)$ obtaining
\begin{equation}
    \boldsymbol{q}(0) = \frac{p^{cr}\,r}{4} 
    \begin{bmatrix}
        8\,\frac{k_r}{k_E}-1-k_r & 0\\
        0 & 7- k_r
    \end{bmatrix},
    \label{eq:q_0}
\end{equation}
which has eigenvalues $\lambda_1=\frac{p^{cr}\,r}{4}\,(8\,\frac{k_r}{k_E}-1-k_r)$ and $\lambda_2=\frac{p^{cr}\,r}{4}\,(7- k_r)$. Then, we define $\lambda_{\min}$ and $\lambda_{\max}$ as the minimum and maximum eigenvalues, respectively.
Accordingly, the eigenvalue ratio is
\begin{equation}
    \frac{\lambda_{\min}}{\lambda_{\max}}=\begin{cases}
        \frac{7-k_r}{8\,\frac{k_r}{k_E}-1-k_r}\quad\text{if}\quad k_E\leq k_r\\[4pt]
        \frac{8\,\frac{k_r}{k_E}-1-k_r}{7-k_r}\quad\text{if}\quad k_E>k_r
    \end{cases}.
\end{equation}
Since we are in 2D, $\boldsymbol{q}(0)$ has rank~2, and the Rayleigh ratio $\mathsf{R}=0$ only when the eigenvalue ratio is negative. Note that the eigenvalue ratio is independent of $r$. Additionally, we can see that for $k_E=k_r$, we obtain $\frac{\lambda_{\min}}{\lambda_{\max}}=1$.

\begin{figure}[H]
    \centering
    \includegraphics[width=1\linewidth]{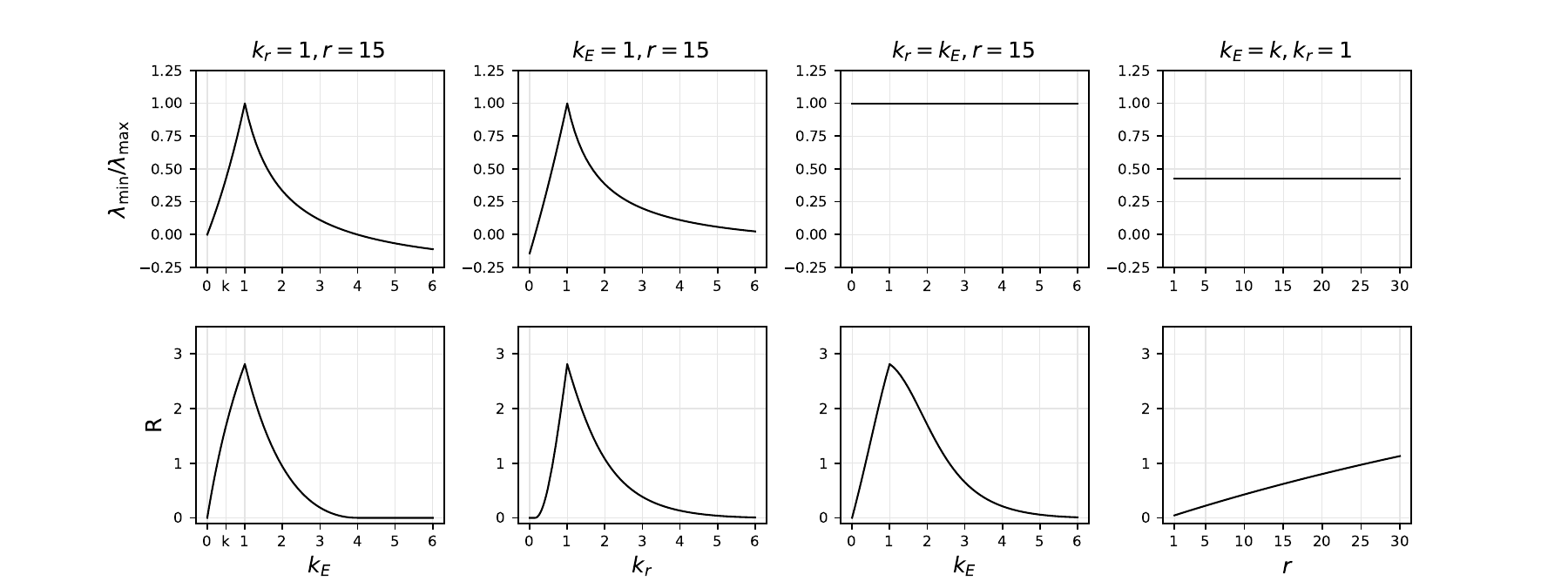}
    \caption{Ratio of the eigenvalues \(\lambda_{\rm max}\), \(\lambda_{\rm min}\) of the tensor \(\boldsymbol{q}(0)\), and Rayleigh ratio \(\mathsf{R}\), for different values of \(k_E\), \(k_r\), and \(r\). The parameter \(p\) is set to $2.0$.
}
    \label{fig:eigen_vol}
\end{figure}

In Figure~\ref{fig:eigen_vol}, we plot the eigenvalue ratio in the first row and the corresponding Rayleigh ratio in the second row for different values of \(k_E\), \(k_r\), and \(r\). In particular, in the first three columns we fix \(r = 15\); note that this affects only the Rayleigh ratio and not the eigenvalue ratio, which is independent of \(r\), as confirmed by the trend observed in the fourth column. 

In the first column of Figure~\ref{fig:eigen_vol}, we examine the case \(k_r = 1\) for different values of \(k_E\). The eigenvalue ratio is non-positive only for \(k_E = 0\) and \(k_E \geq 4\), where the Rayleigh ratio vanishes, indicating that the homogeneous solution is unstable. For \(k_r = 1\), both the eigenvalue and Rayleigh ratios reach a maximum at \(k_E = 1\), corresponding to the isotropic material case. The Rayleigh ratio decreases as \(k_E\) deviates from this value in either direction, showing that anisotropy promotes the loss of stability of the homogeneous solution, i.e., $\mathsf{R}<1$. This is consistent with \cite{zolesi2024stability,vicentini2025variational}, which show that, in a phase-field model of cohesive fracture, states of \(\boldsymbol{q}(0)\) dominated by the volumetric component tend to stabilize the homogeneous solution. Accordingly, deviations from a purely volumetric state, as induced here by anisotropy, favor instability. The same argument also applies to the Rayleigh ratio trends in the second and third columns of Figure~\ref{fig:eigen_vol}, where we again observe that anisotropy promotes the instability of the homogeneous solution.

The increase of the Rayleigh ratio with increasing \(r\) for fixed \(k_E\) and \(k_r\), shown in the fourth column of Figure~\ref{fig:eigen_vol}, is consistent with the observations in \cite{zolesi2024stability, vicentini2025variational}. In these works on phase-field modeling of cohesive fracture, it has also been observed that when \(\boldsymbol{q}(d)\) is neither a rank-one tensor nor a rank-two tensor with non-zero eigenvalues of opposite signs, increasing \(r\)—i.e., the ratio of the cohesive length to the regularization length—favors the stability of the homogeneous solution.

Based on \eqref{eq:p_cr} and \eqref{eq:q_0}, we observe that the components of \(\boldsymbol{q}(0)\) are proportional to \(\sqrt{r}\). Consequently, the components of \(\boldsymbol{\eta}(\boldsymbol{k})\) are also proportional to \(\sqrt{r}\), which implies that the numerator of the Rayleigh ratio in \eqref{eq:Rayleigh_ratio}, i.e., \(\boldsymbol{q}(\bar{d}):\mathbb{C}(\bar{d}):\boldsymbol{q}(\bar{d}) - \mathsf{V}\), is proportional to \(r\). Furthermore, for the case \(k_r = 1\), the denominator of the Rayleigh ratio, \(\tfrac{1}{2}\,\bar{\boldsymbol{\sigma}}:\mathbb{S}''(\bar{d}):\bar{\boldsymbol{\sigma}}\), becomes independent of \(r\). Hence, for \(k_r = 1\), the Rayleigh ratio \(\mathsf{R}\) varies linearly with \(r\), which explains the linear trend observed in Figure~\ref{fig:eigen_vol}. For the general case \(k_r \neq 1\), the trend would be non-linear with \(r\).

\subsection{Uniaxial stress state}
\label{AppB_uni}
\begin{figure}[H]
    \centering
    \includegraphics[width=1\linewidth]{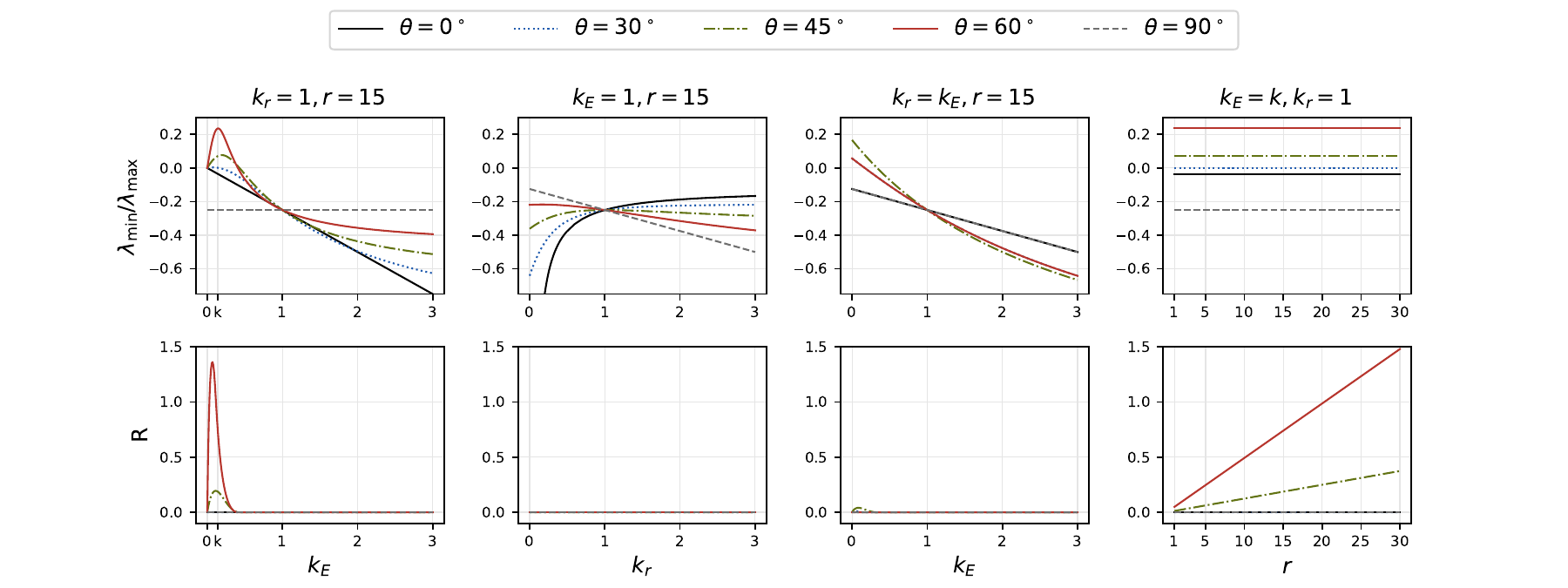}
    \caption{Ratio of the eigenvalues \(\lambda_{max}\), \(\lambda_{min}\) of the tensor \(\boldsymbol{q}(0)\) for different values of $k_E$ and $\theta$. The parameter $p$ is set to $2.0$.}
    \label{fig:eigen_uniaxial}
\end{figure}

With reference to Figure~\ref{Figure:Sketch_BarBVP}, neglecting the out-of-plane direction to obtain a simplified 2D setting, let us consider the \(\theta\)-dependent stress state
\begin{equation}
\bar{\boldsymbol{\sigma}} = \sigma^{cr}(\theta) 
\begin{bmatrix}
\cos^{2}\theta & -\sin\theta\cos\theta \\
-\sin\theta\cos\theta & \sin^{2}\theta
\end{bmatrix}
\end{equation}
which corresponds to the stress \(\tilde{\boldsymbol{\sigma}}\) in \eqref{1D_rotated}, with \(\sigma^{cr}(\theta)\) given by \eqref{eq:CriticalStressThetaMCM}. 
Using the definition in \eqref{eq:Rayleigh_ratio}, we compute \(\boldsymbol{q}(0)\) and, from it, determine \(\lambda_{\rm min}\) and \(\lambda_{\rm max}\), which are the minimum and maximum eigenvalues of \(\boldsymbol{q}(0)\), respectively.  
In Figure~\ref{fig:eigen_uniaxial}, similarly to the case of the volumetric stress state, we plot the eigenvalue ratio in the first row and the corresponding Rayleigh ratio \(\mathsf{R}\) in the second row for different values of \(k_E\), \(k_r\), \(r\), and, in this case, also for different angles \(\theta\).

With reference to the first column of Figure~\ref{fig:eigen_uniaxial}, where \(k_r = 1\) and \(r = 15\), we observe that for uniaxial stress states the deviatoric component of \(\boldsymbol{q}(0)\) dominates over the volumetric one, making the eigenvalue ratio \(\lambda_{\min}/\lambda_{\max}\) predominantly negative and thus leading to a Rayleigh ratio \(\mathsf{R} = 0\). However, due to the intrinsic material anisotropy governed by \(k_E\), it is possible to find loading angles \(\theta\) for which the volumetric contribution of \(\boldsymbol{q}(0)\) increases, resulting in eigenvalues of the same sign and consequently \(\mathsf{R} > 0\).  
In this case, the stability must be assessed by evaluating the Rayleigh ratio \(\mathsf{R}\) and checking whether it is smaller or greater than 1. The figure shows that for a loading angle of \(\theta = 60^\circ\), unstable homogeneous solutions may occur, i.e., \(\mathsf{R} \geq 1\).  
Similar considerations apply to the second and third columns of Figure~\ref{fig:eigen_uniaxial}. In the second column (\(k_E = 1\)), the eigenvalue ratio remains negative, yielding \(\mathsf{R} = 0\). In the third column (\(k_r = k_E\)), there exist combinations of \(k_E\) and \(\theta\) for which the eigenvalue ratio becomes positive; however, also in this case, \(\mathsf{R} < 1\), indicating that the value \(r = 15\) is not sufficiently large to stabilize the homogeneous solution.

Similarly to the case of the volumetric stress state, in the fourth column of Figure~\ref{fig:eigen_uniaxial} (\(k_E =k= 0.1401\), corresponding to the peak of the eigenvalue ratio in the first column, and \(k_r = 1\)), we observe that the eigenvalue ratio is independent of \(r\). In agreement with previous stability analyses of phase-field models for cohesive fracture \cite{zolesi2024stability, vicentini2025variational}, in the same column we observe that the Rayleigh ratio \(\mathsf{R}\) increases with \(r\) when \(\mathsf{R} \neq 0\), meaning that a larger cohesive length relative to the regularization length favors the stability of the homogeneous solution.

\SetKwComment{Comment}{/* }{ */}

\bibliographystyle{ieeetr}
\bibliography{BibliographyPaper1,References}

\end{document}